\documentclass[11pt,draftcls,onecolumn]{IEEEtran}

\usepackage[cmex10]{amsmath}
\usepackage{algorithmic}
\usepackage{array}
\usepackage{mdwmath}
\usepackage{mdwtab}
\usepackage[tight,footnotesize]{subfigure}
\usepackage[font=footnotesize]{subfig}
\usepackage{color}
\usepackage{algorithm}
\usepackage{afterpage}  
\usepackage{float}      
\usepackage{graphicx}   
\usepackage{lscape}     
\usepackage{psfrag}     
\usepackage[cmex10]{amsmath}    
\usepackage{longtable}  
\usepackage{fancyhdr}   
\usepackage{enumerate}  
\usepackage{amssymb}    
\usepackage{setspace}   
\usepackage{makeidx}    
\usepackage{multirow}
\usepackage{cite}
\usepackage{amsmath,amssymb,amsthm,color}
\usepackage{amsmath,amssymb,color}
\usepackage{graphicx,psfrag,url,lineno}
\usepackage{enumerate}

\newcommand{\Z}{\mathbb{Z}}

\newcommand{\R}{\mathbb{R}}

\newtheorem{thm}{Theorem}[]

\newtheorem{prop}[thm]{Proposition}

\theoremstyle{remark}
\newtheorem{rem}{Remark}[thm]
\theoremstyle{definition}
\newtheorem{defn}[thm]{Definition}

\newcommand{\supp}{\text{supp}}
\def\argmin{\mathop{\rm arg\,min}}
\def\minim{\mathop{\hbox{minimize}}}
\theoremstyle{definition}
\newtheorem*{rem*}{Remark}

\begin{document}
%
\title{Recovering Compressively Sampled Signals Using Partial Support Information}
\author{
  Michael P. Friedlander\thanks{This work was supported in part by the Natural Sciences and Engineering Research Council of Canada (NSERC) Collaborative Research and Development Grant DNOISE II (375142-08).}\thanks{Michael P. Friedlander is with the Department of Computer Science, The University of British  Columbia, Vancouver, Canada.}, 
  Hassan Mansour*\thanks{Hassan Mansour is with the Departments of Computer Science and Mathematics, The University of British  Columbia, Vancouver, Canada.}, 
  Rayan Saab\thanks{Rayan Saab is with the Department of Mathematics, Duke University, Durham, USA.}, 
  and {\"O}zg{\"u}r Y{\i}lmaz \thanks{{\"O}zg{\"u}r Y{\i}lmaz  is with the Department of Mathematics, The University of British  Columbia, Vancouver, Canada.}}

\maketitle
\begin{abstract}
  We study recovery conditions of weighted $\ell_1$ minimization for
  signal reconstruction from compressed sensing measurements when
  partial support information is available. We show that if at least
  $50\%$ of the (partial) support information is accurate, then
  weighted $\ell_1$ minimization is stable and robust under weaker
  sufficient conditions than the analogous conditions for standard
  $\ell_1$ minimization. Moreover, weighted $\ell_1$ minimization
  provides better upper bounds on the reconstruction error in terms of
  the measurement noise and the compressibility of the signal to be
  recovered.  We illustrate our results with extensive numerical
  experiments on synthetic data and real audio and video signals.
\end{abstract}



\IEEEpeerreviewmaketitle
\section{Introduction}

Compressed sensing (see, e.g., \cite{Donoho2006_CS, CRT05, CRT06}) is
a paradigm for effective acquisition of signals that admit sparse (or
approximately sparse) representations in some transform domain. The
approach can be used to reliably recover such signals from
significantly fewer linear measurements than their ambient
dimension. Because a wide range of natural and man-made
signals---e.g., audio, natural and seismic images, video, and wideband
radio frequency signals---are sparse or approximately sparse in
appropriate transform domains, the potential applications of
compressed sensing can be immense.

Let $\Sigma_k^N:=\{x\in \R^N : \|x\|_0 \leq k\}$ be the set of
all $k$-sparse signals in $\R^N$, and let 
\begin{equation}y := Ax + e
\label{eq:measurements}
\end{equation} 
be a vector of measurements where $A$ is a known $n\times N$
measurement matrix, and $e$ denotes additive noise that satisfies
$\|e\|_2\leq \epsilon$ for some known $\epsilon\ge 0$. Compressed
sensing theory states that it is possible to recover $x\in\Sigma_k^N$
from $y$ (given $A$) even when $n \ll N$, i.e., using very few
measurements. For example, when $e=0$, one may recover an estimate
$x^*$ of the signal $x$ as the solution of the constrained $\ell_0$
minimization problem
\begin{equation}\label{eq:L0_min}
\minim_{z\in \R^N}\ \|z\|_0 \ \text{subject to} \ \ Az=y.
\end{equation}
In fact, using \eqref{eq:L0_min}, any $x\in\Sigma_k^N$ can be
recovered perfectly using $n$ measurements when $n > 2k$ and $A$ is in
general position (see, e.g., \cite{Donoho03}). However, $\ell_0$
minimization is a combinatorial problem and quickly becomes
intractable as the dimensions increase. Instead, the convex
relaxation
\begin{equation}\label{eq:L1_min}
 \minim_{z \in \R^N}\ \|z\|_1 \ \text{subject to} \ \|Az - y\|_2 \leq \epsilon
\end{equation}
can be used to recover the estimate $x^*$.  Cand{\'e}s, Romberg and
Tao \cite{CRT05} and Donoho \cite{Donoho2006_CS} show that if $n
\gtrsim k\log(N/k)$, then $\ell_1$ minimization \eqref{eq:L1_min}
can stably and robustly recover $x$ from ``incomplete'' and inaccurate
measurements $y = Ax + e$, where $A$ is an appropriately chosen $n
\times N$ measurement matrix and $\|e\|_2 \leq \epsilon$. Note that
compressed sensing is a \emph{non-adaptive} data acquisition technique
because the measurement matrix $A$ does not depend on $x$, the signal
being measured. Furthermore, the recovery method that we just
described is itself non-adaptive because no information on $x$ is used
in \eqref{eq:L1_min}.  Our goal in this paper is to examine a
recovery method that is adaptive in the sense that it exploits prior
support information on $x$; the measurement process, however, remains
non-adaptive.



\subsection{Compressed sensing with prior support information}
The $\ell_1$ minimization problem \eqref{eq:L1_min} does not
incorporate any prior information about the support of $x$. However,
in many applications it may be possible to
draw an estimate of the support of the signal or an estimate of its
largest coefficients. For example, signals such as video and audio exhibit
correlation over temporal frames that can be exploited to
estimate a portion of the support using previously decoded frames.

Consider the example where $x \in \R^N$ is a compressible signal,
i.e., it can be well-approximated by its $k$ largest-in-magnitude
entries, where $k \ll N$. If $x$ represents the discrete cosine
transform (DCT) or wavelet coefficients of an image, then the entries
of $x$ that correspond to the low frequency subbands are most
likely to be non-zero and carry most of the energy of the signal
\cite{Robucci_etal_CS_CMOS:2010}. In such cases, it is beneficial to
incorporate this information in the recovery algorithm when $x$ is
compressively sampled. 

\subsection{Previous Work}

We are especially interested in methods that incorporate prior support
information by replacing the $\ell_1$ minimization in
\eqref{eq:L1_min} with weighted $\ell_1$ minimization
\begin{equation}\label{eq:weighted_L1_0}
  \minim_{z}\ \|z\|_{1,\mathrm{w}}\ \text{subject to}\ \|Az - y\|_2 \leq \epsilon,
\end{equation}
where $\mathrm{w}\in [0,1]^N$ and $\|z\|_{1,\mathrm{w}} := \sum_i
\mathrm{w}_i |z_i|$ is the weighted $\ell_1$ norm. In particular, in
the methods that we describe here (including our own proposed method),
the main idea is to choose $\mathrm{w}$ such that the entries of $x$
that are ``expected'' to be large are penalized less in this weighted
objective function.

The recovery of compressively sampled signals using prior support
information has been previously studied in the literature; see, e.g., 
\cite{CS_using_PI_Borries:2007, Vaswani_Lu_Modified-CS:2010,
  Vaswani_ISIT:2009, reg_mod_BPDN_Vaswani:2010, Jacques:2010,
  WL1_min_Hassibi:2009}. In fact, the problem of sparse recovery with partially known support was independently introduced in three  works -- in von Borries et al. \cite{CS_using_PI_Borries:2007}, in Vaswani and Lu \cite{Vaswani_ISIT:2009}; and in Khajehnejad et al. \cite{ WL1_min_Hassibi:2009}. 
  
 The work by Borries et
al. \cite{CS_using_PI_Borries:2007} demonstrated empirically that
incorporating support information of a signal with a sparse discrete
Fourier transform (DFT) allows for the number of compressed sensing
measurements to be reduced by exactly the size of the known part of
the support. 
 Borries et al. achieve this by using a weighted $\ell_1$
minimization approach with zero weights on the known support.

More recently, Vaswani and Lu \cite{Vaswani_Lu_Modified-CS:2010,
  Vaswani_ISIT:2009, reg_mod_BPDN_Vaswani:2010} proposed a modified
compressed sensing approach that again incorporates known support
elements using a weighted $\ell_1$ minimization approach with zero
weights on the known support. Their work derives sufficient recovery
conditions for the noise free case (i.e., set $e=0$ in
\eqref{eq:measurements} and $\epsilon=0$ in \eqref{eq:weighted_L1_0})
that are weaker than the analogous $\ell_1$ minimization conditions of
\cite{CRT05} in the case where a large proportion of the support is
known. This work is supplemented by a regularized modified compressed
sensing approach that deals with noisy measurements
\cite{reg_mod_BPDN_Vaswani:2010}. The work of Vaswani and Lu was also
extended by Jacques in \cite{Jacques:2010} to the cases of
compressible signals and noisy measurements. The approach of Jacques
is based on studying the \textit{innovative} basis pursuit denoising
(\textit{i}BPDN) problem, which minimizes weighted $\ell_1$-norm of
the solution with with zero weights applied to the support estimate;
Jacques and shows that (\textit{i}BPDN) has a similar stability
behavior to the unweighted $\ell_1$ problem.

A similar method is proposed by Khajehnejad et
al.~\cite{WL1_min_Hassibi:2009} for the recovery of compressively
sampled signals with support information. The performance of this
method is analyzed using a Grassman angle approach. Prior information
is defined in terms of two disjoint sets that partition
$\{1,\ldots,N\}$. The elements in the first set have a probability
$P_1$ of being nonzero, and the elements in the second set have a
probability $P_2$ of being nonzero, where $P_1 \neq P_2$. The authors
propose weighted $\ell_1$ minimization to recover the unknown vector
where different weights $w_1$ and $w_2$ are assigned to the elements
in the two sets. In particular, they find the class of signals $x$,
depending on $P_j$ and $w_j$, $j=1,2$, which can be recovered with
high probability using weighted $\ell_1$.

Finally, the weighted $\ell_1$ minimization problem is related to the
``adaptive lasso" described the statistics literature and studied by
Zou in \cite{Zou:2006}; it is defined by
$$
 \minim\limits_{z}\ \|y - Az\|_2^2 + \lambda_n \sum\limits_{j = 1}^{N} w_j|z_j|,
$$ 
where $\lambda_n$ varies with the sample size $n$ such that
$\lambda_n/\sqrt{n} \rightarrow 0$ and $\lambda_{n}n^{(\gamma-1)/2}
\rightarrow \infty$ for some $\gamma > 0$, and the weights $w_j =
1/|\hat{x}_j|^{\gamma}$, where $\hat{x}$ is given signal estimate
that is root-$n$ consistent\footnote{Root-$n$ consistency means
  that if $\hat{x}^*$ is the solution to the adaptive lasso problem,
  then $\sqrt{n}(\hat{x}^* - x) \rightarrow \mathcal{N}(0, \sigma^2)$
  in distribution, where $\sigma$ depends on the noise variance and
  the covariance of the measurement matrix $A$.}. However, the problem
studied by Zou addresses the overdetermined scenario where the ambient
dimension $N$ of the signal is fixed and the number of measurements $n
\rightarrow \infty$. In this case, Zou shows that the adaptive
lasso enjoys the oracle properties but acknowledges that when $N > n
\rightarrow \infty$ it is nontrivial to find a consistent estimate for
constructing the weights in the adaptive lasso.

\subsection{Contributions}

In this paper we adopt the weighted $\ell_1$ minimization approach
described by \eqref{eq:weighted_L1_0}. Given a support estimate $\widetilde{T}
\subset \{1,2,\dots,N\}$ for $x$, we set $w_j=\omega \in [0,1]$ whenever
$j\in \widetilde{T}$, and $w_j=1$ otherwise. Unlike Borries et al. or Vaswani et
al., in our results we allow $\omega$ to be non-zero. We derive
stability and robustness guarantees for weighted $\ell_1$ minimization
that generalize the results of \cite{CRT05}. Our results take into
consideration the accuracy of the support estimate. In particular, we
prove that if the (partial) support estimate is at least 50\%
accurate, then weighted $\ell_1$ minimization outperforms standard
$\ell_1$ minimization in terms of accuracy, stability, and
robustness. Finally, we note that when $\omega=0$, our results hold
under weaker sufficient conditions than those in
\cite{Vaswani_Lu_Modified-CS:2010}.

In Section~\ref{sec:CSoverview}, we review the $\ell_1$ recovery
guarantees of \cite{CRT05}. In Section~\ref{sec:WeightedL1}, we state our main
result and compare our theoretical
results with standard $\ell_1$ recovery as well as the results of
\cite{Vaswani_Lu_Modified-CS:2010, Vaswani_ISIT:2009}. In Sections~\ref{sec:Simulations} and \ref{sec:applications}, we present
the outcome of numerical experiments on synthetic and
on audio and video signals. We conclude with the proof of our main
theorem in Section~\ref{sec:Proof}.


\section{Compressed Sensing Overview}\label{sec:CSoverview}

Consider an arbitrary signal $x \in \R^N$ and let $x_k \in \Sigma_k^N$
be its best $k$-term approximation. Let $T_0 = \supp(x_k)$,
where $T_0\subseteq \{1,\ldots,N\}$ and $|T_0| \leq k$. We wish to
reconstruct the signal $x$ from $y = Ax + e$, where $A$ is a known
$n\times N$ measurement matrix with $n \ll N$, and $e$ denotes the
(unknown) measurement error that satisfies $\|e\|_2 \leq \epsilon$ for
some known margin $\epsilon>0$.

As we mentioned in the introduction, it was shown in \cite{CRT05} that
$x$ can be stably and robustly recovered from the measurements $y$ by
solving the optimization problem \eqref{eq:L0_min} if the measurement
matrix $A$ has the {\em restricted isometry property} (RIP), also
defined by \cite{CRT05}.
\begin{defn}\label{def:RIP}
  The restricted
  isometry constant $\delta_k$ of a matrix $A$ is the smallest number
  such that for all $k$-sparse vectors $u \in \Sigma_k^N$,
	\begin{equation}\label{eq:RIP}
		(1-\delta_k)\|u\|_2^2 \leq \|Au\|_2^2 \leq (1+\delta_k)\|u\|_2^2.
	\end{equation} 
\end{defn}
\noindent Cand\`es et al.\cite{CRT05} use the RIP to provide
conditions and bounds for stable and robust recovery of $x$ by solving
(\ref{eq:L1_min}).
\begin{thm}[Cand$\grave{\textrm{e}}$s, Romberg, Tao \cite{CRT05}]\label{thm:L1_recovery}
  Suppose that $x$ is an arbitrary vector in $\R^N$, and let $x_k$ be
  the best $k$-term approximation of $x$. Suppose that there exists an
  $a \in \frac{1}{k}\Z$ with $a>1$ and
  \begin{equation}\label{eq:suff_L1}
    \delta_{ak} + a\delta_{(1+a)k} < a-1.
  \end{equation}
  Then the solution $x^*$ to
  (\ref{eq:L1_min}) obeys
	\begin{equation}\label{eq:L1_recovery}
          \|x^*-x\|_2 \leq C_0\epsilon + C_1 k^{-1/2}\|x - x_k\|_1.
	\end{equation}
\end{thm}

\begin{rem}
The constants in Theorem \ref{thm:L1_recovery} are explicitly given by
\large{
\begin{equation}\label{eq:L1_constants}
\begin{array}{ll}
	C_0 = \frac{2\left(1+a^{-1/2}\right)}{\sqrt{1-\delta_{(a+1)k}} - a^{-1/2}\sqrt{1+\delta_{ak}}},& C_1 = \frac{2 a^{-1/2}\left(\sqrt{1-\delta_{(a+1)k}} + \sqrt{1+\delta_{ak}}\right) }{\sqrt{1-\delta_{(a+1)k}} - a^{-1/2}\sqrt{1+\delta_{ak}}}.
\end{array}
\end{equation}
}
\normalsize
\end{rem}

From Theorem \ref{thm:L1_recovery}, one can see that if $A$ satisfies
(the slightly stronger condition)
\begin{equation}\label{eq:L1_RIP}
	\delta_{(a+1)k} < \frac{a-1}{a+1},
\end{equation}
then the constrained $\ell_1$ minimization problem in
(\ref{eq:L1_min}) recovers $x$ with an approximation error that scales
well with measurement noise and the ``compressibility'' of $x$.
Moreover, if $x$ is sufficiently sparse (i.e., $x = x_k$), and the
measurement process is noise-free, then Theorem \ref{thm:L1_recovery}
guarantees exact recovery of $x$ from $y$.

{\section{Compressed sensing with partial support estimation}\label{sec:WeightedL1}

  In this section, we present our main result showing that weighted
  $\ell_1$ minimization can be used to stably and robustly recover
  sparse and compressible signals from noisy measurements when there
  is partial (and possibly partly inaccurate) prior support
  information. Our result holds under weaker sufficient conditions
  than its counterpart for $\ell_1$ minimization when the support
  estimate is more than $50\%$ accurate. Moreover, it results in
  smaller error bounds. We also compare our results with the modified
  compressed sensing approach proposed in
  \cite{Vaswani_Lu_Modified-CS:2010}.

\subsection{Weighted $\ell_1$ minimization with estimated support}
Let $T_0$ be the support of $x_k$, and let $\widetilde{T}$, the
support estimate, be a
subset of $\{1,2,\ldots,N\}$ with cardinality $k_1 := |\widetilde{T}|
= \rho k$, where $0 \leq \rho \leq a$ for some $a > 1$. 
As before, we wish to recover an arbitrary vector $x \in \R^N$ from
noisy compressive measurements $y = Ax + e$, where $e$ satisfies
$\|e\|_2 \leq \epsilon$. To recover $x \in \R^N$, we now consider the
weighted $\ell_1$ minimization problem with the following choice of weights:
\begin{equation}\label{eq:weighted_L1}
  \minim_{z}\ \|z\|_{1,\mathrm{w}}\ \text{subject to}\ \|Az - y\|_2 \leq \epsilon
  \quad\text{with}\quad
  \mathrm{w}_i = 
  \begin{cases}
      1,      & i \in \widetilde{T}^c,
    \\\omega, & i \in \widetilde{T}. 
  \end{cases}
\end{equation}
Here, $0 \leq \omega \leq 1$ and $\|z\|_{1,\mathrm{w}}$ is as defined
in (\ref{eq:weighted_L1_0}). Our main result follows.

\begin{thm}\label{thm:weighted_L1_recovery}
  Let $x$ be in $\R^N$ and let $x_k$ be its best $k$-term
  approximation, supported on $T_0$.  Let
  $\widetilde{T}\subset\{1,\ldots,N\}$ be an arbitrary set and define
  $\rho$ and $\alpha$ as before such that $|\widetilde{T}|=\rho k$ and
  $|\widetilde{T} \cap T_0| = \alpha\rho k$.  Suppose that there
  exists an $a\in \frac{1}{k}\Z$, with $a \geq (1-\alpha)\rho$, $a>1$,
  and the measurement matrix $A$ has RIP with
\begin{equation}\label{eq:suff_wl1}
\delta_{ak} + \frac{a}{\left(\omega + (1-\omega)\sqrt{1+\rho-2\alpha\rho}\right)^2}\delta_{(a+1)k} < \frac{a}{\left(\omega + (1-\omega)\sqrt{1+\rho-2\alpha\rho}\right)^2} - 1, 
\end{equation}
for some given $0 \leq \omega \leq 1$. 
 Then the solution $x^*$ to (\ref{eq:weighted_L1}) obeys
\begin{equation}\label{eq:weighted_L1_recovery}
	\|x^* - x\|_2 \leq C_0'\epsilon + C_1'k^{-1/2}\left(\omega\|x - x_k\|_1 + (1-\omega)\|x_{\widetilde{T}^c\cap T_0^c}\|_1\right),
\end{equation}
where $C_0'$ and $C_1'$ are well-behaved constants that depend on the measurement matrix $A$, the weight $\omega$, and the parameters $\alpha$ and $\rho$.
\end{thm}
 The proof of the theorem is presented in section \ref{sec:Proof}.
 \begin{rem} Note that the parameters in
   Theorem~\ref{thm:weighted_L1_recovery} specify two important
   ratios: $\rho$ determines the ratio of the size of the estimated
   support to the size of the actual support of $x_k$ (or the support
   of $x$ if $x$ is $k$-sparse). On the other hand, $\alpha$ determines
   the ratio of the number of indices in $\supp(x_k)$ that were
   accurately estimated in $\widetilde{T}$ to the size of
   $\widetilde{T}$. Specifically, $\alpha = \frac{|\widetilde{T}\cap T_0|}{|\widetilde{T}|}$.
\end{rem}
\begin{rem}\label{rem:const}
The constants $C_0'$ and $C_1'$ are explicitly given by the expressions
\begin{equation}\label{eq:weighted_L1_constants}
\begin{array}{ll}
	C_0' = \frac{\textstyle 2\left(1+\frac{\omega +
              (1-\omega)\sqrt{1+\rho-2\alpha\rho}}{\sqrt{a}}\right)}{\textstyle\sqrt{1-\delta_{(a+1)k}}
          - \frac{\omega +
            (1-\omega)\sqrt{1+\rho-2\alpha\rho}}{\sqrt{a}}\sqrt{1+\delta_{ak}}},&
        C_1' = \frac{\textstyle 2
          a^{-1/2}\left(\sqrt{1-\delta_{(a+1)k}} +
            \sqrt{1+\delta_{ak}}\right) }{\textstyle \sqrt{1-\delta_{(a+1)k}} - \frac{\omega + (1-\omega)\sqrt{1+\rho-2\alpha\rho}}{\sqrt{a}}\sqrt{1+\delta_{ak}}}.
\end{array}
\end{equation}

\noindent Consequently, Theorem~\ref{thm:weighted_L1_recovery}, with
$\omega=1$, reduces to the stable and robust recovery theorem of
\cite{CRT05}, which we stated above---see
Theorem~\ref{thm:L1_recovery}.
\end{rem}

\begin{rem} It is sufficient that $A$ satisfies
\begin{equation}\label{eq:weighted_L1_RIP}
	\delta_{(a+1)k} < \hat{\delta}^{(\omega)} := \frac{a - \left(\omega + (1-\omega)\sqrt{1+\rho - 2\alpha\rho}\right)^2}{a + \left(\omega + (1-\omega)\sqrt{1+\rho - 			2\alpha\rho}\right)^2}
\end{equation}
for Theorem \ref{thm:weighted_L1_recovery} to hold, i.e., to guarantee
stable and robust recovery of the signal $x$ from measurements $y = Ax
+ e$ (with constants $C_0'$ and $C_1'$ given in
\eqref{eq:weighted_L1_constants} and \eqref{eq:weighted_L1_RIP}).
\end{rem}

\begin{rem}
  Theorems \ref{thm:L1_recovery} and \ref{thm:weighted_L1_recovery}
  guarantee stable and robust recovery for matrices $A$ satisfying a
  condition on $\delta_{(a+1)k}$ with $a > 1$. A slightly different
  approach was used by Cand{\`e}s \cite{candes2008rip} to handle the
  case $a=1$. Cand{\`e}s proved that if $\delta_{2k} <
  (\sqrt{2}+1)^{-1}$, then $\ell_1$ minimization (\ref{eq:L1_min})
  achieves stable and robust recovery. Following the same technique,
  with appropriate modifications to handle the weighted $\ell_1$
  objective, we can derive the analogous alternative sufficient
  condition
\begin{equation}\label{eq:weighted_2k_RIP}
\delta_{2k} < \left(\sqrt{2}(\omega + (1-\omega)\sqrt{1+\rho - 2\alpha\rho}) + 1\right)^{-1},
\end{equation}
which guarantees stable and robust recovery using weighted $\ell_1$
minimization (\ref{eq:weighted_L1}). We omit the details of this
calculation.
\end{rem}  

\subsection{Comparison to standard $\ell_1$ recovery}\label{sec:Compare_with_L1}
In this section, we compare the sufficient conditions for
Theorem~\ref{thm:weighted_L1_recovery} and
Theorem~\ref{thm:L1_recovery} as well as the associated constants of
these two theorems. The following observation is easy to verify.
\begin{prop} Let $C_0,C_1,C_0'$, and $C_1'$ be as above. Then
\begin{enumerate}[(i)]
\item If $\omega=1$, then $C_0'=C_0$, $C_1'=C_1$, and the sufficient
  conditions for Theorem~\ref{thm:weighted_L1_recovery}, given in
  \eqref{eq:suff_wl1}, are identical to those of
  Theorem~\ref{thm:L1_recovery}, given in \eqref{eq:suff_L1}. 
\item If $\alpha=0.5$, then, again $C_0'=C_0$, $C_1'=C_1$, and the sufficient
  conditions for Theorem~\ref{thm:weighted_L1_recovery}, given in
  \eqref{eq:suff_wl1}, are identical to those of
  Theorem~\ref{thm:L1_recovery}, given in \eqref{eq:suff_L1}.
\item Suppose $0\le \omega <1$. Then $C_0'< C_0$ and $C_1'< C_1$
  if and only if $\alpha > 0.5$.  
\end{enumerate}
\end{prop}


\begin{figure*}[t]
	\centering
	
	\subfigure[]{\includegraphics[width=3.2in]{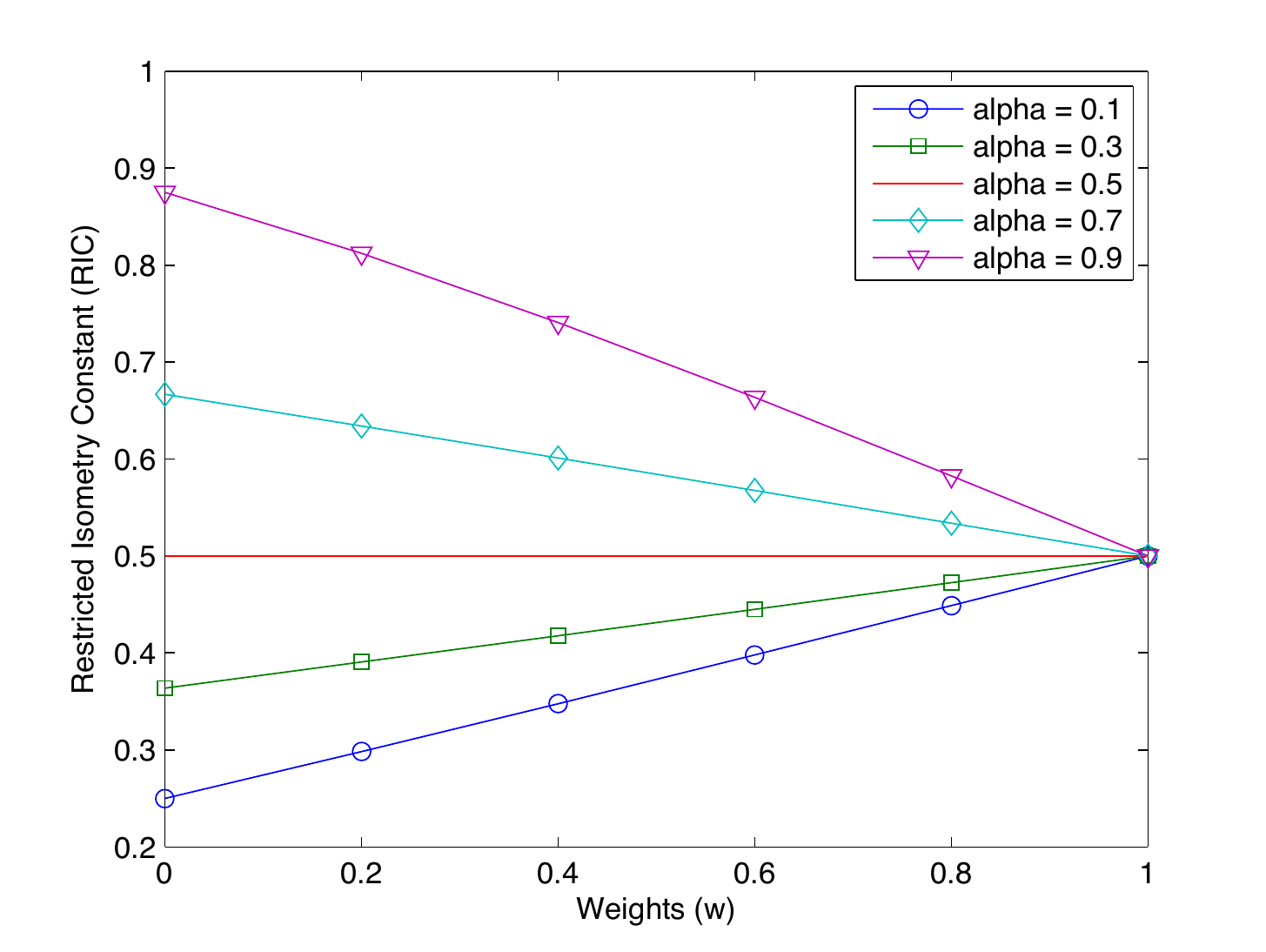}	}
	\mbox{
	\subfigure[]{\includegraphics[width=3.2in]{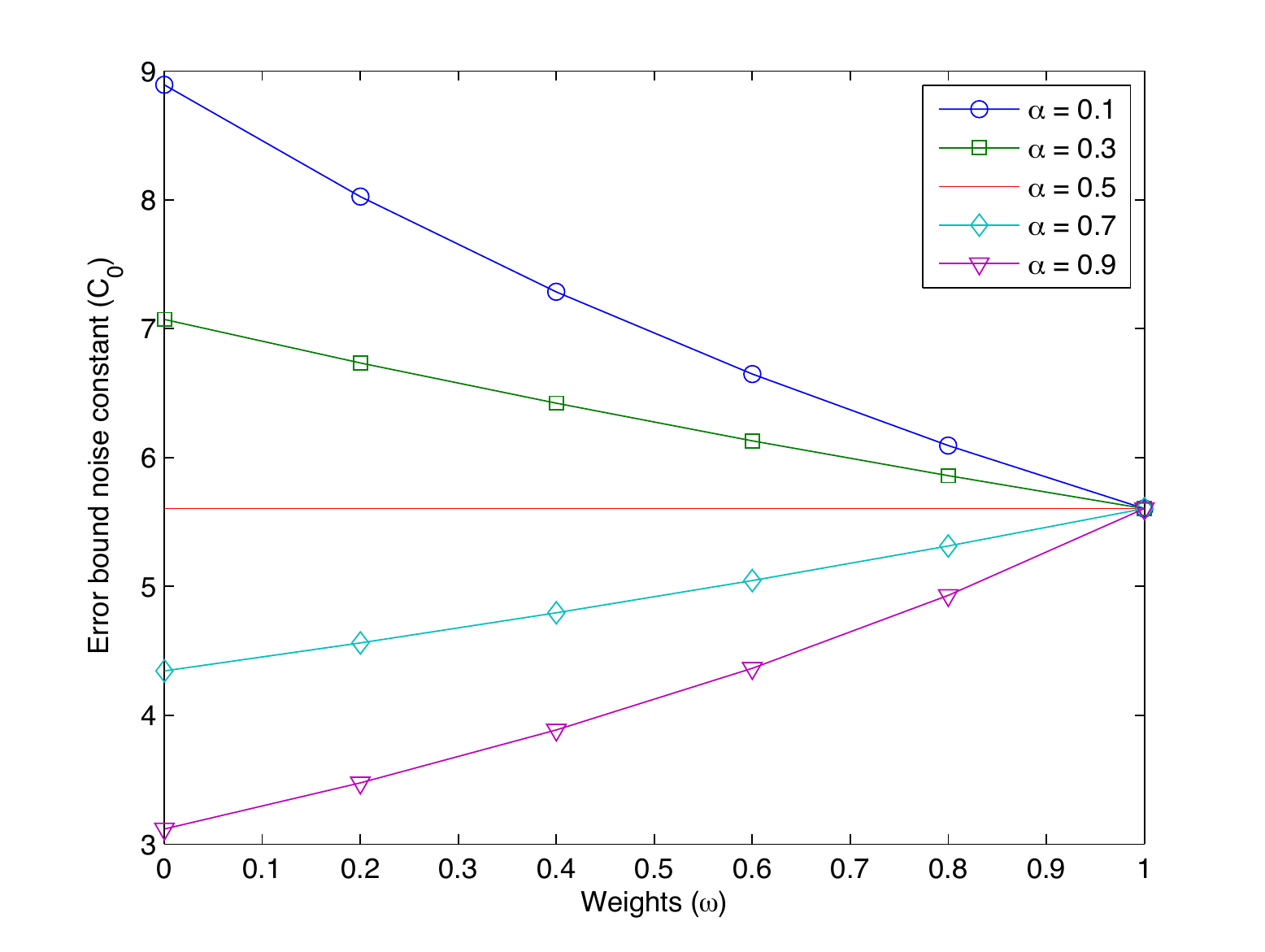}}
	\subfigure[]{\includegraphics[width=3.2in]{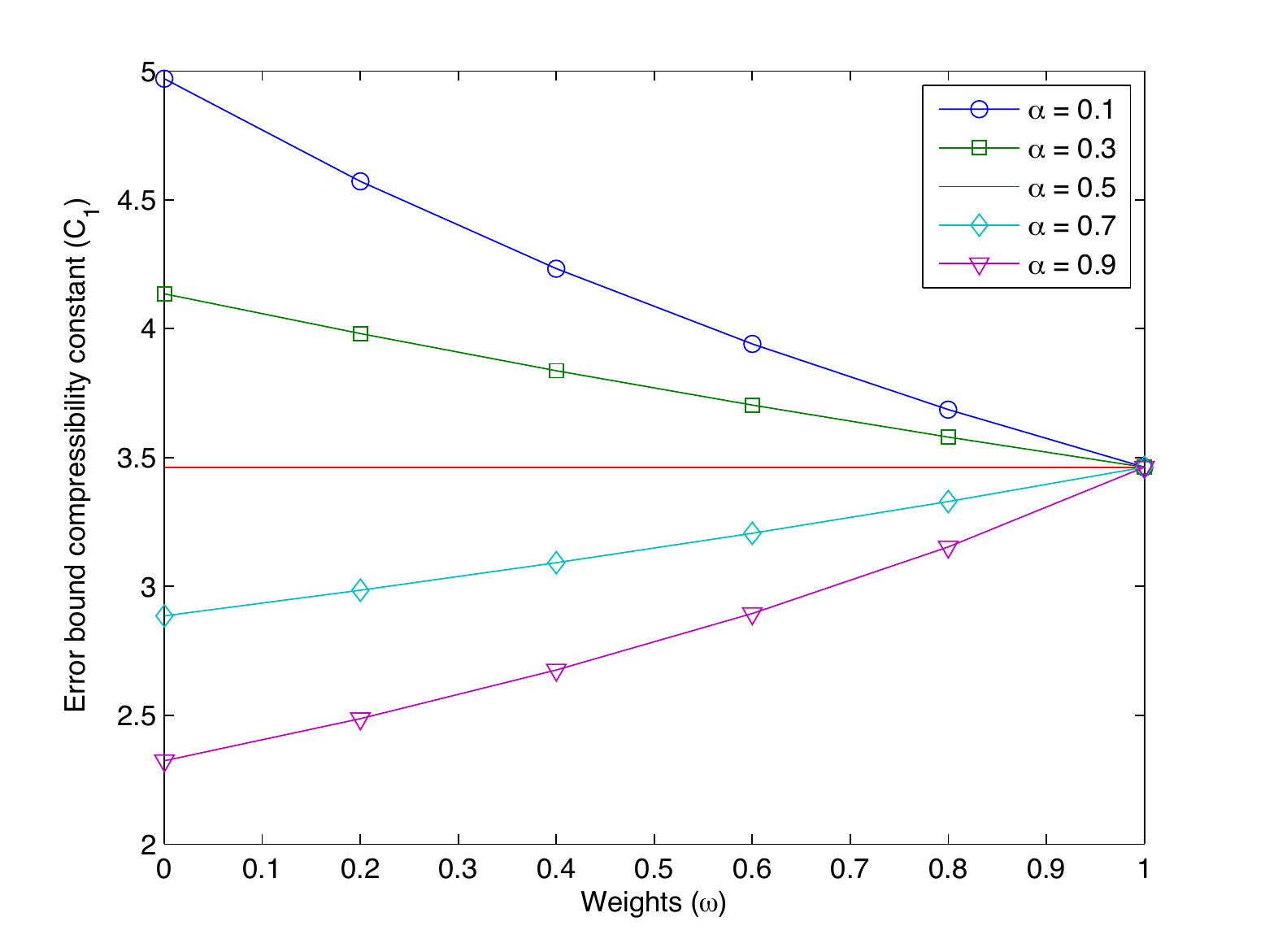}}
	}
	\caption{Comparison of the sufficient conditions for recovery
          and stability constants for weighted $\ell_1$
          reconstruction with various of $\alpha$. In all the figures,
          we set $a=3$ and $\rho=1$. (a) $\hat{\delta}^{(\omega)}$
          vs. $\omega$, (b) $C_0'$ vs. $\omega$, (c) $C_1'$
          vs. $\omega$. In (b) and (c) we fix $\delta_{(a+1)k} = 0.1$.}
	\label{fig:RIP_C0_C1_v_weights}
\end{figure*}

Next, we illustrate how the slightly stronger sufficient conditions
given in \eqref{eq:weighted_L1_RIP} and the respective stability
constants vary with $\alpha$ and $\omega$. Recall that when
$\omega=1$, \eqref{eq:weighted_L1_RIP} reduces to
\eqref{eq:L1_RIP}. In Figure~\ref{fig:RIP_C0_C1_v_weights} (a), we
plot, for different values of $\alpha$, $\hat{\delta}^{(\omega)}$ as
defined in \eqref{eq:weighted_L1_RIP}, versus $\omega$, where we set
the parameter $a=3$. We observe that as $\alpha$ increases the
sufficient condition on the RIP constant becomes weaker, allowing for
a wider class of measurement matrices $A$. For example, with $a=3$,
when $70\%$ of the support estimate is accurate, with $\omega=0.2$ it
suffices to have $\hat\delta^{(\omega)}<0.763$, compared with
$\hat\delta^{(1)}<0.5$ for $\ell_1$ minimization. Figures
\ref{fig:RIP_C0_C1_v_weights} (b) and (c) illustrate that for a fixed
matrix $A$, the constants $C_0'$ and $C_1'$ decrease as $\alpha$
increases. 
Note that compared to setting $\omega=0$, assigning non-zero weights
$\omega$ adds robustness to the weighted $\ell_1$ problem in the case
when $\alpha < 0.5$, i.e., when we have an inaccurate support estimate
$\widetilde{T}$ with more than half the entries falling outside the
support of the best $k$-term approximation of $x$. This could be
beneficial in applications where the accuracy of the support estimates
vary significantly from one signal to the next. Furthermore, in
numerical experiments (see Section~\ref{sec:Simulations}) we observe
that using non-zero weights improves the quality of the
reconstruction, especially in the noisy and compressible settings, not
only when $\alpha < 0.5$ but also in some cases where $\alpha>0.5$. A
mathematical understanding of this behavior and of how to optimally
choose the weight $\omega$ is beyond the scope of this paper.

\subsection{The zero weight case: $\omega=0$}

One special case of the weighted $\ell_1$ problem that is of interest
is the zero weight case, i.e., set $\omega = 0$ in
\eqref{eq:weighted_L1}. It can be seen from
Figure \ref{fig:RIP_C0_C1_v_weights} that recovery using weighted
$\ell_1$ minimization (\ref{eq:weighted_L1}) achieves the smallest
error bound constants at $\omega = 0$ when $\alpha > 0.5$. On the
other hand, the recovery performance is worst when $\omega = 0$ and
$\alpha < 0.5$, i.e., when the support estimate is highly inaccurate.

Several contributions in the literature adopt the zero-weight approach,
mainly in applications where prior support information is assumed to
be highly accurate, i.e., $\alpha$ is close to 1, e.g., see
\cite{CS_using_PI_Borries:2007, Vaswani_Lu_Modified-CS:2010,
  modified_BPDN_Vaswani:2010}. The most recent study to address this
problem is the work by Vaswani and Lu
\cite{Vaswani_Lu_Modified-CS:2010} where a sufficient condition in
terms of the RIP of the matrix $A$ is derived for exact recovery in
the noise free case. Another work by the same authors
\cite{modified_BPDN_Vaswani:2010} addresses the noisy case, however,
the recovery algorithm in this case is different from
\eqref{eq:weighted_L1} in that the objective function is modified to
include a regularization term. The sufficient condition derived in
Corollary 1 of \cite{Vaswani_Lu_Modified-CS:2010} is expressed as
\begin{equation}\label{eq:VaswaniRIC}
 2\delta_{2u} + \delta_{3u} + \delta_{k} + \delta^2_{k+u} + 2\delta^2_{k + 2u} < 1,
\end{equation}
where $u = (1-\alpha\rho)k$ is the size of the unkown support. Recall that $\alpha$ is such that $\alpha\rho k$ is the size of the known support.

Below we compare our condition \eqref{eq:weighted_2k_RIP} with that of
\cite{Vaswani_Lu_Modified-CS:2010} given in \eqref{eq:VaswaniRIC} for
different values of the unknown support size $u$. We consider the case
$\rho = 1$, $u > 0$, and $n/N = 0.5$. Thus, \eqref{eq:weighted_2k_RIP}
reduces to 
\begin{equation} \label{eq:red_suff_u}
\delta_{2k} < \frac{1}{2\sqrt{u/k} + 1}. 
\end{equation}
Since the two sufficient
conditions, i.e., \eqref{eq:red_suff_u} and \eqref{eq:VaswaniRIC}, are expressed in terms of RIP constants of different-sized
submatrices of $A$, a simple comparison of the upper bounds is not
informative. For this reeason, we restrict our attention to
measurement matrices drawn from the Gaussian ensemble and we estimate
the associated RIP constants (i.e., $\delta_{2k},\delta_{2u},\dots$)
for such matrices using the bounds derived in \cite{Bah_Tanner_RIPbounds:2010}.
In particular, we calculate the ratios $u/k$ that satisfy the
conditions \eqref{eq:VaswaniRIC} and \eqref{eq:red_suff_u},
respectively, and plot the results in Figure
\ref{fig:uk_vs_kn}. Observe that for the same measurement matrix $A$
and sparsity level $k/n$, our sufficient condition guarantees the
recovery of $k$-sparse signals with significantly less accurate prior
support information compared to the condition of Vaswani et
al. \cite{Vaswani_Lu_Modified-CS:2010}.
\begin{figure}[t]
	\centering
	\includegraphics[width=4.5in]{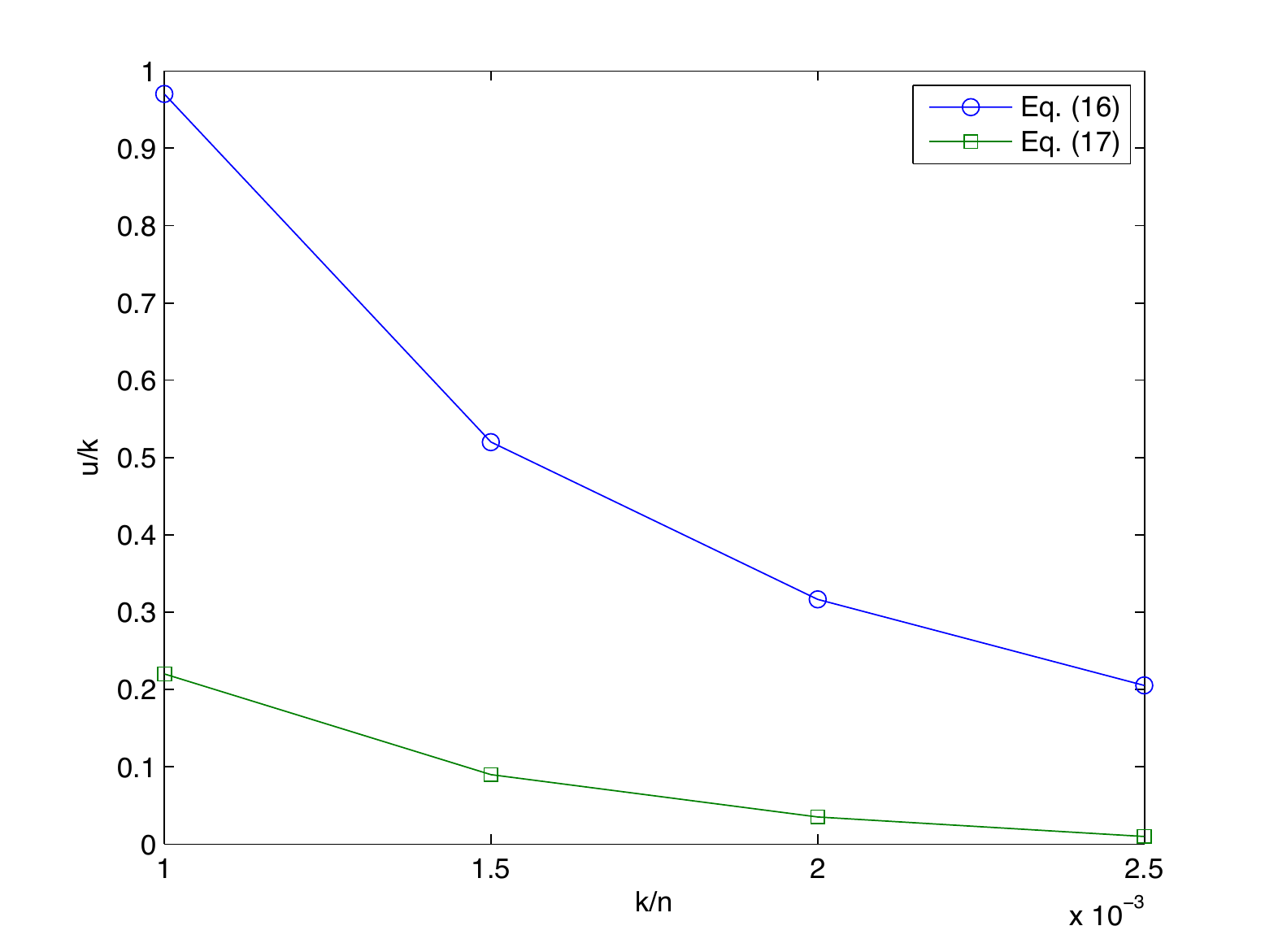}
	\caption{Comparison between the values of $u/k$ that satisfy
          each of the sufficient conditions \eqref{eq:VaswaniRIC} and
          \eqref{eq:red_suff_u}. The measurement matrix has Gaussian
          entries with $n/N = 0.5$.}
	\label{fig:uk_vs_kn}
\end{figure} 

It is clear in Figure~\ref{fig:uk_vs_kn} that our recovery
guarantees are superior to those of \cite{Vaswani_Lu_Modified-CS:2010}
at least when the aspect ratio of the measurement matrix is
$n/N=0.5$. Next, we shall focus on cases where we have a highly
accurate estimate of the full support of the $k$-sparse vector $x$. In
other words, we set $\rho=1$ as above and consider values of $\alpha$
that are close to 1. For these cases, we will compare our theoretical
guarantees to those of \cite{Vaswani_Lu_Modified-CS:2010} for various
values of the measurement matrix aspect ratio. To that end, we observe
that the left-hand side of \eqref{eq:VaswaniRIC} is increasing in
$u$. Thus, for any $u$, \eqref{eq:VaswaniRIC} can hold only if
$$
\delta_{k} + 3\delta_{k}^2 < 1 \implies \delta_k < 0.4343, 
$$
which is obtained by setting $u=0$ in \eqref{eq:VaswaniRIC} and
observing that $\delta_0=0$ by definition. On the other hand, using
the bounds from \cite{Bah_Tanner_RIPbounds:2010}, we can estimate
$\delta_{2k}$ and find the corresponding range of $u$ for
\eqref{eq:red_suff_u} to hold in the case when $A$ is a Gaussian
random matrix. The upper bound on the range of $u/k$ for various aspect
ratios of the measurement matrix is reported in
Table~\ref{tab:RIPrecoveryGuarantee}. We conclude that in various
cases with different measurement matrix aspect ratios our theoretical
results guarantee recovery while the results of
\cite{Vaswani_Lu_Modified-CS:2010} fail to provide any recovery
guarantee.

\begin{table}[t]
	\centering
		\begin{tabular}{|c|c|c|c|c|}
		\hline
		\textbf{n/N} & \textbf{k/n} & $\mathbf{\delta_k}$ & $\mathbf{\delta_{2k}}$ & \textbf{u/k} \\ \hline
		0.1 & 0.0029 & 0.4343 & 0.6153 & 0.0978 \\ \hline
		0.2 & 0.0031 & 0.4343 & 0.6139 & 0.0989 \\ \hline	
		0.3 & 0.003218 & 0.4343 & 0.61176 & 0.1007 \\ \hline
		0.4 & 0.003315 & 0.4343 & 0.61077 & 0.1015 \\ \hline
		0.5 & 0.003394 & 0.4343 & 0.60989 & 0.1023 \\ \hline
		\end{tabular}
	\caption{Maximum unknown support size $u/k$ for which
         \eqref{eq:red_suff_u} holds while \eqref{eq:VaswaniRIC}
          fails to hold. For a given aspect ratio $n/N$, we compute
          the value of $k/n$ for which $\delta_k=0.4343$. This value,
          using \eqref{eq:red_suff_u}, yields the corresponding bound on $u/k$.}
	\label{tab:RIPrecoveryGuarantee}
\end{table}


We finish this section by comparing the recovery guarantees we obtain
in the zero-weight case with conditions that guarantee recovery via
$\ell_1$ minimization without using any prior support information. To
this end, we present the phase diagrams of measurement matrices $A$
with Gaussian entries that satisfy the conditions on the restricted
isometry constants $\delta_{(a+1)k}$ given in \eqref{eq:L1_RIP} and
\eqref{eq:weighted_L1_RIP} with $\omega = 0$, respectively. We use the
bounds derived in \cite{Bah_Tanner_RIPbounds:2010} and plot the curves
in Figure \ref{fig:phase_diagrams} for matrices satisfying the
sufficient conditions on $\delta_{4k}$ with $\rho = 1$ and $\alpha =
0.3$, $0.6$, and $0.8$.
\begin{figure}[t]
	\centering
	\includegraphics[width=4.5in]{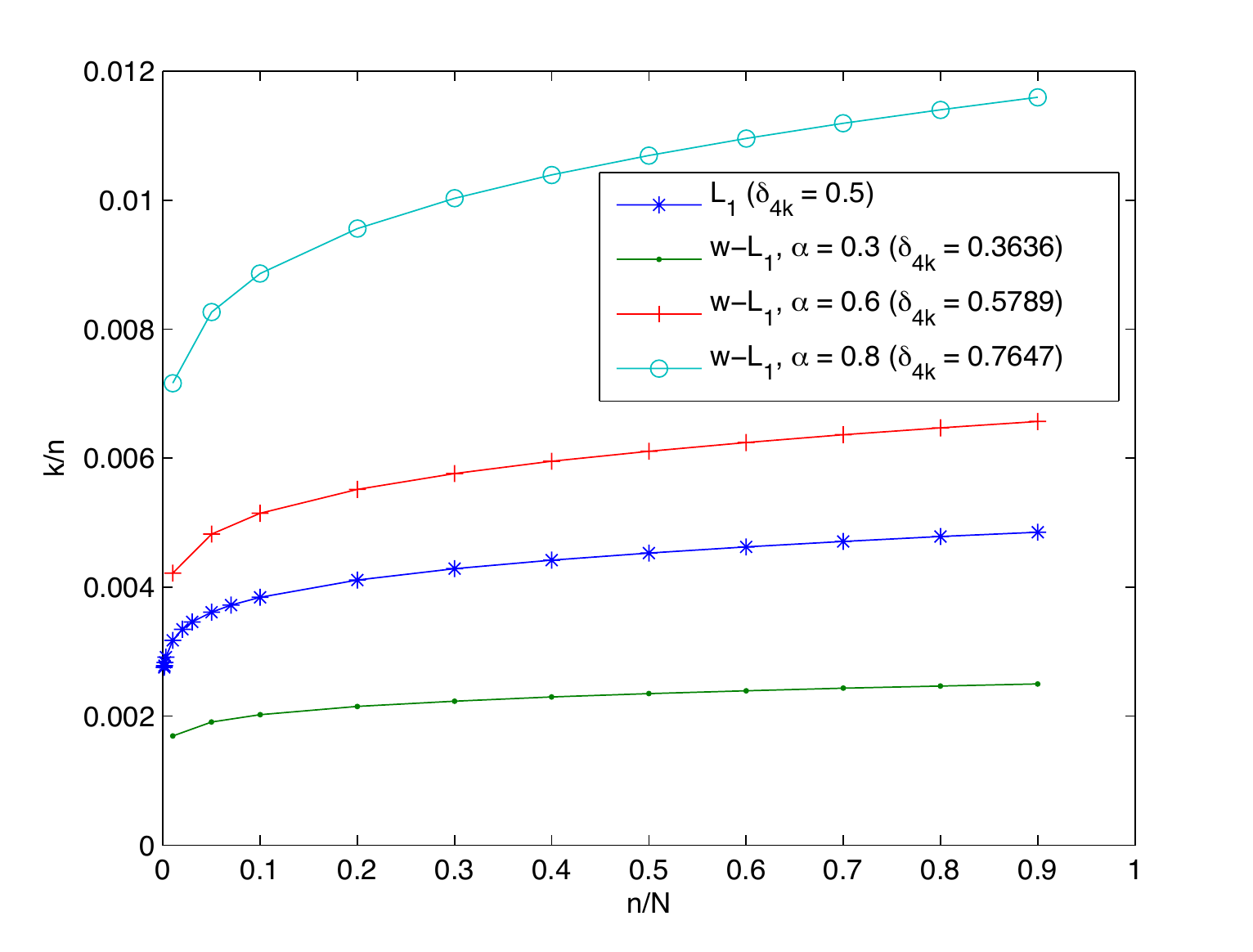}
	\caption{Comparison between the phase diagrams of measurement matrices with Gaussian entries satisfying the sufficient recovery conditions of standard $\ell_1$ minimization and weighted $\ell_1$ minimization with $\omega = 0$ and $\alpha = 0.3$, 0.6, and 0.8. The plots are calculated using the upper bounds on the restricted isometry constants derived in \cite{Bah_Tanner_RIPbounds:2010}.}
	\label{fig:phase_diagrams}
\end{figure}

\section{Numerical Examples}\label{sec:Simulations}
In this section, we present numerical experiments that illustrate the
benefits of using weighted $\ell_1$ minimization to recover sparse and
compressible signals when partial prior support information (which is
possibly inaccurate) is available. To that end, we compare the
recovery capabilities of standard $\ell_1$ and weighted $\ell_1$
minimization for a suite of synthetically generated sparse and
compressible signals. In all of our experiments, we use
SPGL1~\cite{BergFriedlander:2008, spgl1:2007} to solve the standard
and weighted $\ell_1$ minimization problems.

\subsection{The sparse case}
We first generate signals $x$ with an ambient dimension $N = 500$ and
fixed sparsity $k = 40$. We compute the (noisy) compressed measurements of $x$
using a Gaussian random measurement matrix $A$ with dimensions $n \times
N$ where we vary $n$ between 80 and 200 with an increment of
20. In the experiments where the measurements are noisy, we set
$\epsilon=\|x\|_2/20$.

Figure \ref{fig:sparse_SNR_vs_n} shows the average reconstruction signal
to noise ratio (SNR) over 20 experiments when using weighted $\ell_1$
minimization depending on the number of measurements, both in the noise-free and
noisy cases. The SNR is
measured in dB and is given by
\begin{equation}
	\mathrm{SNR}(x,x^*) = 10\log_{10}\left(\frac{\|x\|_2^2}{\|x - x^*\|_2^2}\right),
\end{equation}
where $x$ is the true signal and $x^*$ is the recovered signal. The
recovery is done via \eqref{eq:weighted_L1} using a support estimate
of size $|\widetilde{T}| = 40$ (i.e., $\rho = 1$) where
\begin{itemize}
\item the accuracy $\alpha$ of the support estimate ranges between
  zero and 1,
\item the constant weight $\omega$ ranges between zero and 1 (recall
  that when $\omega = 1$ \eqref{eq:weighted_L1} is equivalent to
  standard $\ell_1$ minimization).
\end{itemize}

Figure \ref{fig:sparse_SNR_vs_n} (a) illustrates that in the noise
free case, the experimental results are consistent with the
theoretical bounds derived in Theorem
\ref{thm:weighted_L1_recovery}. More specifically, it can be seen that
when $\alpha \geq 0.5$ the best recovery is achieved for a weight
$\omega = 0$ whereas a $\omega = 1$ results in the worst SNR. On the
other hand, when $\alpha < 0.5$ the performance of the recovery
algorithms is shifted towards larger values of $\omega$ in the
severely underdetermined cases (small $n$). Figure \ref{fig:sparse_SNR_vs_rho} shows the average recovered SNR using weighted $\ell_1$ minimization for different values of the parameter $\rho$. It is evident from the figure that using a larger support estimate favours better reconstruction, However, it can be seen in both the noise free and noisy measurement vector cases that the recovery is more sensitive to the accuracy $\alpha$ of the support estimate than its size relative to $k$.

\begin{rem}
Recall from Section~\ref{sec:Compare_with_L1}---see Figure
\ref{fig:RIP_C0_C1_v_weights}---that when $x$ is sparse and $\alpha
\geq 0.5$, $\omega = 0$ results in the smallest error bound
constants. Otherwise, i.e., when $\alpha < 0.5$, $\omega = 1$
minimizes the error constants. However, this does not match entirely
with our experimental observations. It can be seen from Figure
\ref{fig:sparse_SNR_vs_n} (b) that, in general, the best recovery is
obtained for intermediate values of $\omega$.

To explain this behaviour, consider the case where the measurement
matrix does not satisfy the RIP conditions for the full recovery of a
$k$-sparse $x$ via weighted $\ell_1$ minimization. In such cases, $x$
can be regarded as compressible: Fix $\hat{k} < k$ be such that
Theorem \ref{thm:weighted_L1_recovery} holds for all $\hat{k}$-sparse
signals and for all $\omega\in [0,1]$. Suppose $\widehat{T}$ is the
support of the best $\hat{k}$ term approximation of $x$. Then Theorem
\ref{thm:weighted_L1_recovery} guarantees stable and robust recovery
of $x$ where the recovery error is bounded by
\begin{equation}\nonumber
	\|x^* - x\|_2 \leq \frac{C_1'(\omega)}{\sqrt{\hat{k}}}\left( \omega\|x_{\widehat{T}^c}\|_1 + (1-\omega)\|x_{\widetilde{T}^c\cap\widehat{T}^c}\|_1\right),
\end{equation}
where $\widetilde{T}$ is the prior support estimate. Denote by
$\hat{\alpha} =
\frac{|\widehat{T}\cap\widetilde{T}|}{|\widetilde{T}|}$ and note that
since $\widehat{T} \subset T_0$, then $\hat{\alpha} < \alpha$.
Focusing our attention on the case when $\hat{\alpha} < 0.5$ (where it
is observed that $0 < \omega < 1$ results in the best recovery), we
make the following observations:
\begin{enumerate}[(i)]
\item The constant $C_1'$ in the error bound above increases as $\omega$
  goes to zero (see Figure \ref{fig:RIP_C0_C1_v_weights}).
\item Since $\widehat{T}^c\cap\widetilde{T}^c \subseteq
  \widehat{T}^c$, the term $\omega\|x_{\widehat{T}^c}\|_1 +
  (1-\omega)\|x_{\widetilde{T}^c\cap\widehat{T}^c}\|_1$ decreases as
  $\omega$ goes to zero.
\end{enumerate}
Therefore, for a fixed $\hat{k}$, there exists $0 \leq \omega \leq 1$ that
minimizes the product of the constant $C_1'$ and the term
$\omega\|x_{\widehat{T}^c}\|_1 +
(1-\omega)\|x_{\widetilde{T}^c\cap\widehat{T}^c}\|_1$. Consequently,
when the algorithm cannot recover the full support of $x$, an
intermediate value of $\omega$ in $[0,1]$ may result in the smallest
recovery error. A full mathematical analysis of the above observations
needs to take into account all the interdependencies between $\omega$,
$\hat{k}$, $\hat{\alpha}$ as well as the parameters in
Theorem~\ref{thm:weighted_L1_recovery} and is beyond the scope of this
paper.
\end{rem}

\begin{figure*}
\centering
\subfigure[Noise Free]{\includegraphics[width=7in]{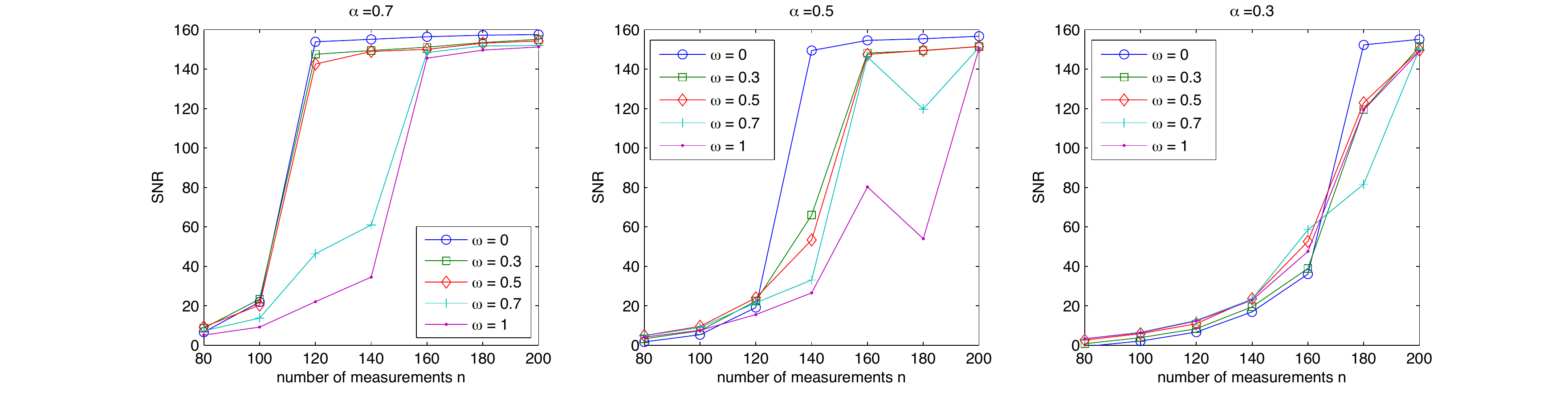}}
\subfigure[5\% Noise Variance]{\includegraphics[width=7in]{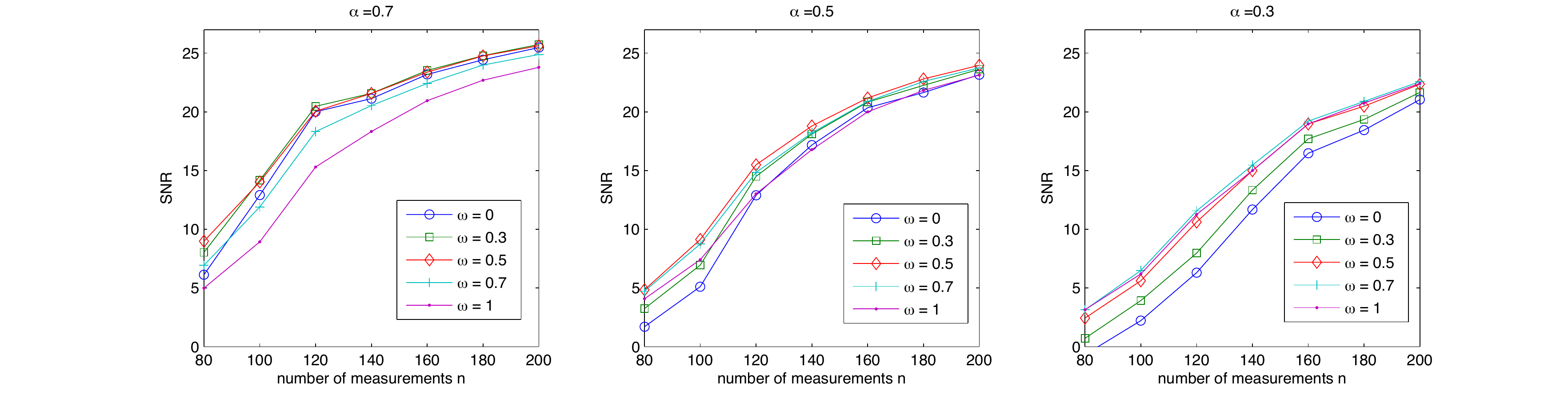}}
\caption{Performance of weighted $\ell_1$ recovery in terms of SNR
  averaged over 20 experiments for sparse signals $x$ with $k = 40$,
  $N = 500$, while varying the number of measurements $n$. From
  left to right, $\alpha = 0.7$, $\alpha = 0.5$, and $\alpha =
  0.3$.}
\label{fig:sparse_SNR_vs_n}
\end{figure*}

\begin{figure*}
\centering
\subfigure[Noise Free]{\includegraphics[width=7in, height=2in]{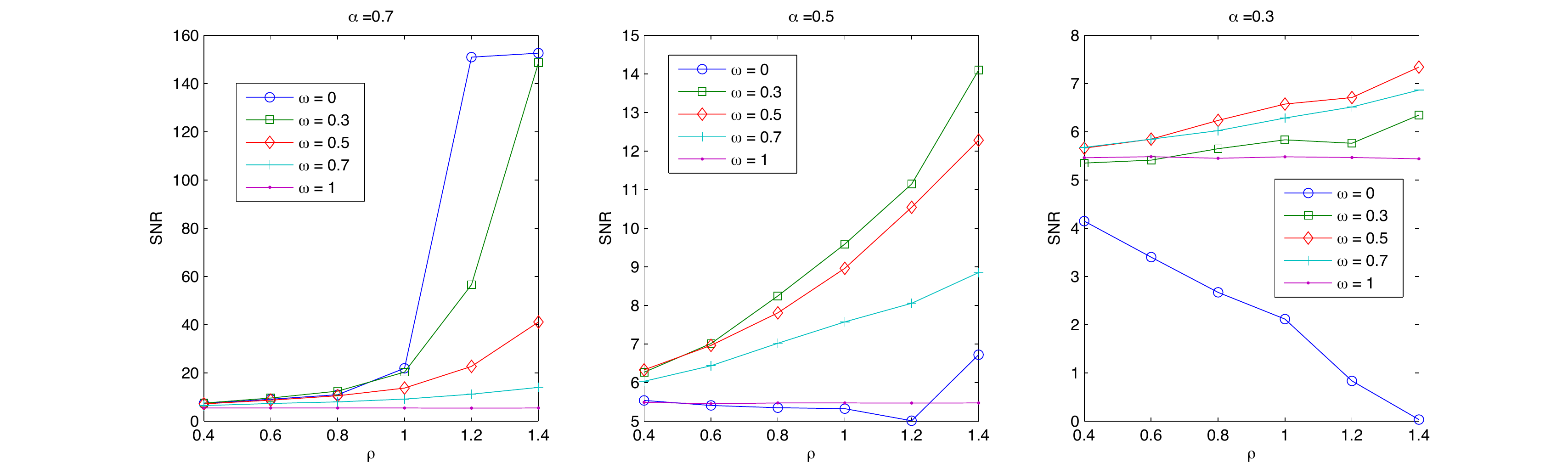}}
\subfigure[5\% Noise Variance]{\includegraphics[width=7in, height=2in]{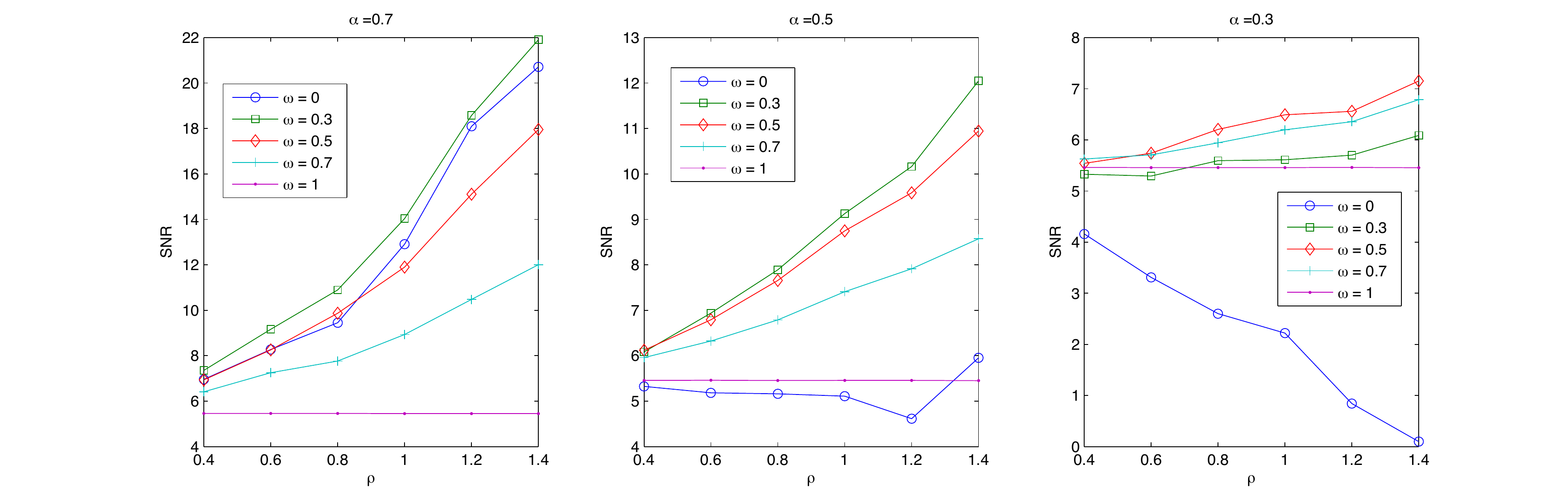}}
\caption{Performance of weighted $\ell_1$ recovery in terms of SNR
  averaged over 20 experiments for sparse signals $x$ with $k = 40$,
  $N = 500$, $n = 100$ while varying the size of the support estimate $\rho$ as a proportion of $k$. From
  left to right, $\alpha = 0.7$, $\alpha = 0.5$, and $\alpha =
  0.3$.}
\label{fig:sparse_SNR_vs_rho}
\end{figure*}

%

\subsection{The compressible case}
Next, we generate a signal $x$ whose coefficients decay like $j^{-p}$
where $j \in \{1,\ldots,N\}$ and $p > 1$. In Figure
\ref{fig:compressible_SNR_vs_rhok_p1.1}, we illustrate the recovered
signal SNR versus the size of the support estimate for $p=1.1$. To
calculate $\alpha$ we set $k=40$, i.e., we are interested in the best
40-term approximation. Notice that on average, a weight $\omega
\approx 0.5$ results in the best recovery. This behavior is consistent
with the explanation provided above where an intermediate value of
$\omega$ balances the tradeoff between the error bound constants and
the norm of the off-support components. We repeat this experiment with
$p=1.5$, $k=20$ and $p=2$, $k=10$. The results are reported in Figures
\ref{fig:compressible_SNR_vs_rhok_p1.5} and
\ref{fig:compressible_SNR_vs_rhok_p2}, and show the same qualitative
behaviour.  

\begin{figure*}
\centering
	\subfigure[Noise free]{\includegraphics[width=7in]{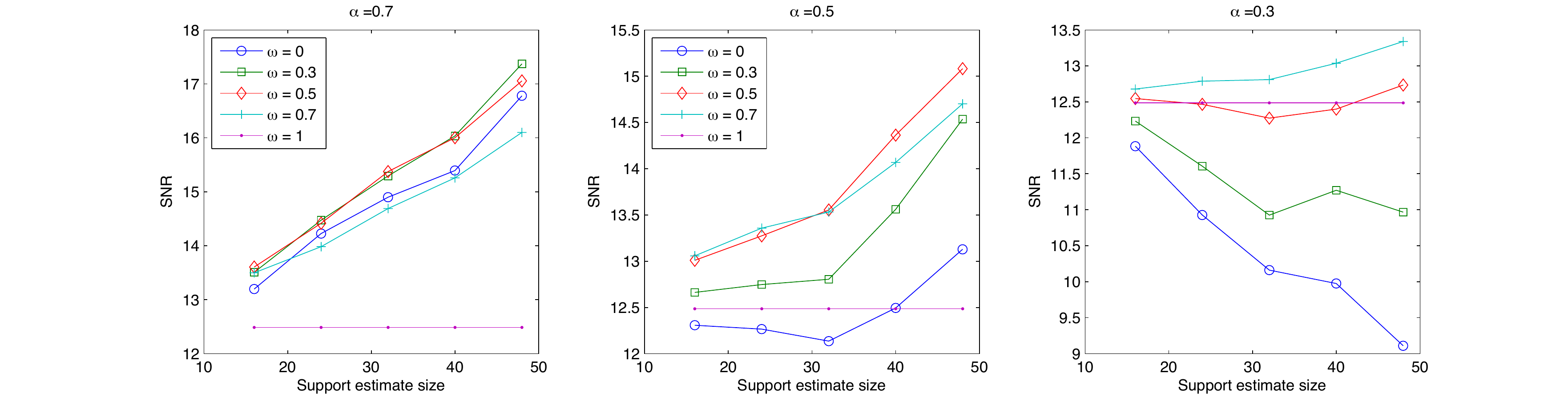}}
	\subfigure[10\% Noise Variance]{\includegraphics[width=7in]{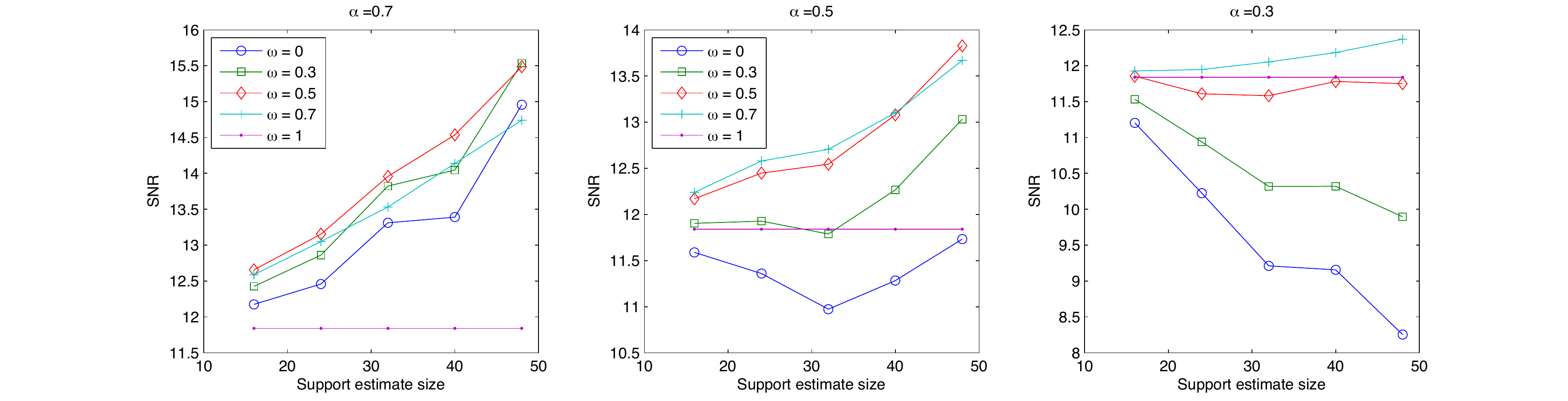}}
	\caption{Performance of weighted $\ell_1$ recovery in terms of SNR averaged over 10 experiments for compressible signals $x$ with $n = 100$, $N = 500$. The coefficients decay with a power $p = 1.1$. The accuracy of the support estimate $\alpha$ is calculated with respect to the best $k = 40$ term approximation. From left to right, $\alpha = 0.7$, $\alpha = 0.5$, and $\alpha = 0.3$.}
	\label{fig:compressible_SNR_vs_rhok_p1.1}
\end{figure*}

\begin{figure*}
\centering
	\subfigure[Noise free]{\includegraphics[width=7in]{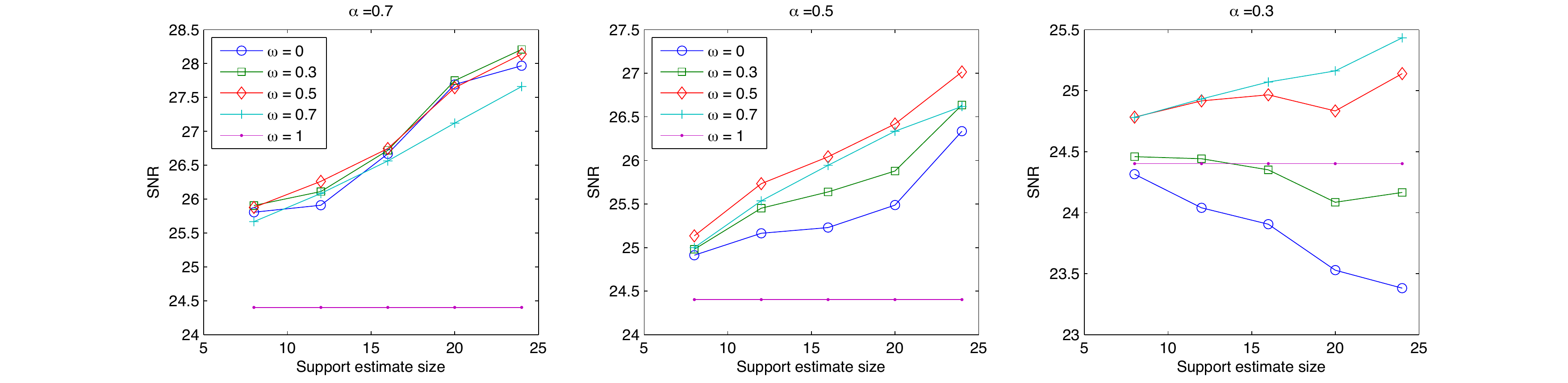}}
	\subfigure[10\% Noise Variance]{\includegraphics[width=7in]{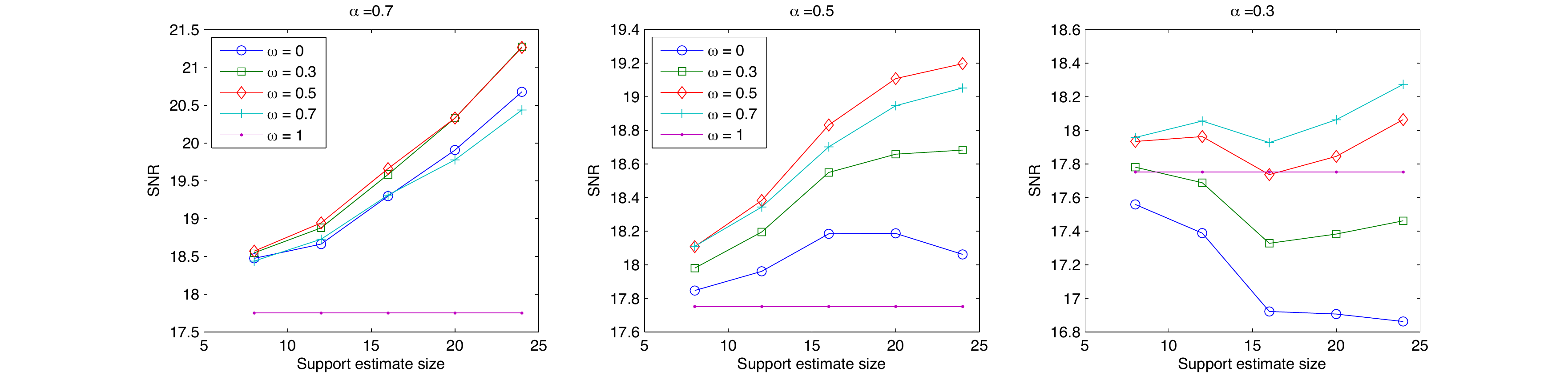}}
	\caption{Performance of weighted $\ell_1$ recovery in terms of SNR averaged over 10 experiments for compressible signals $x$ with $n = 100$, $N = 500$. The coefficients decay with a power $p = 1.5$. The accuracy of the support estimate $\alpha$ is calculated with respect to the best $k = 20$ term approximation. From left to right, $\alpha = 0.7$, $\alpha = 0.5$, and $\alpha = 0.3$.}
	\label{fig:compressible_SNR_vs_rhok_p1.5}
\end{figure*}

\begin{figure*}
\centering
	\subfigure[Noise free]{\includegraphics[width=7in]{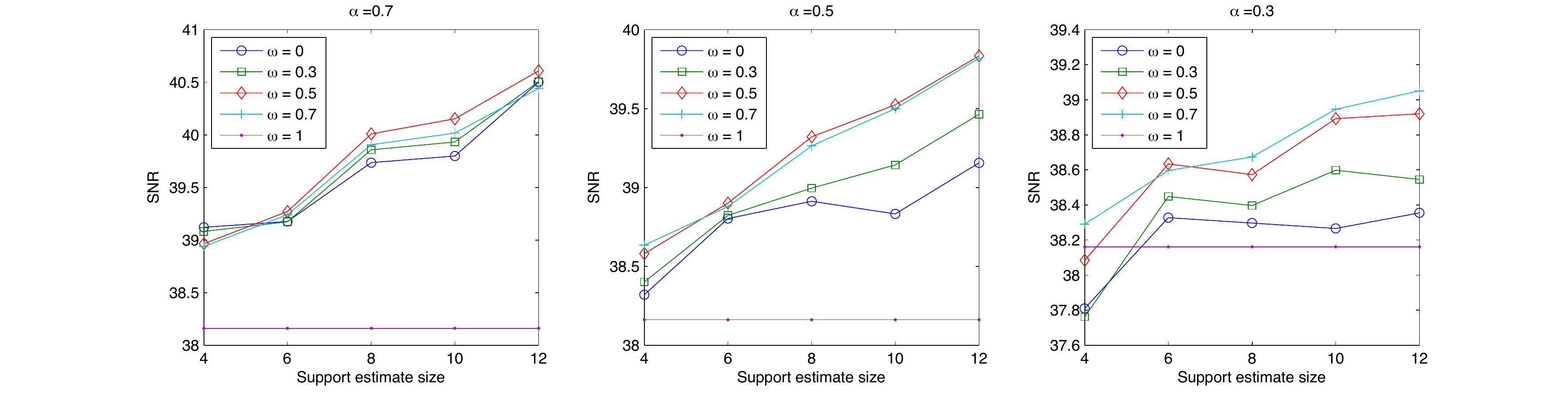}}
	\subfigure[10\% Noise Variance]{\includegraphics[width=7in]{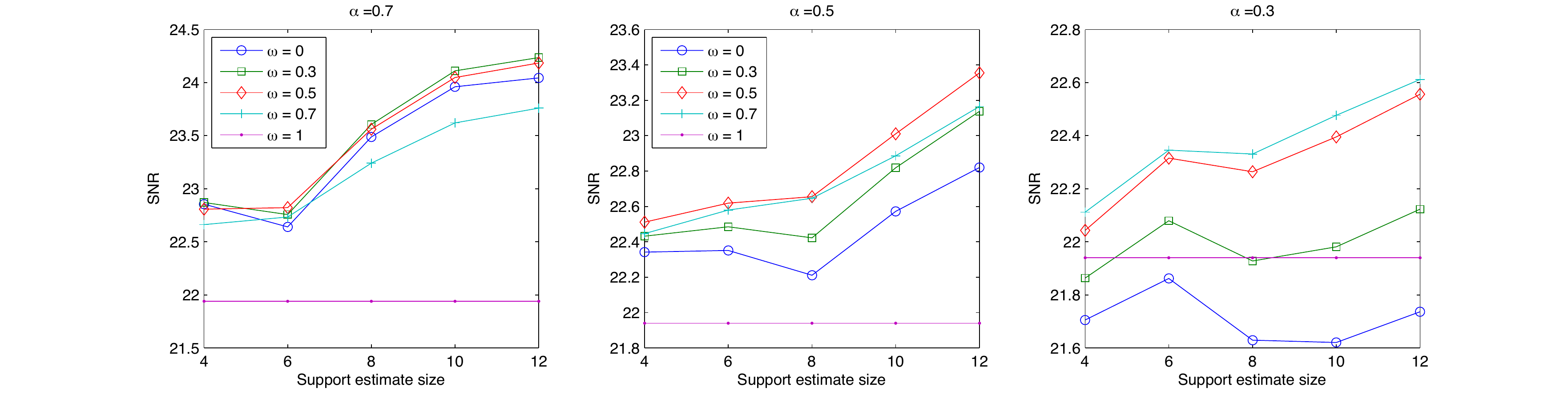}}
	\caption{Performance of weighted $\ell_1$ recovery in terms of SNR averaged over 10 experiments for compressible signals $x$ with $n = 100$, $N = 500$. The coefficients decay with a power $p = 2$. The accuracy of the support estimate $\alpha$ is calculated with respect to the best $k = 10$ term approximation. From left to right, $\alpha = 0.7$, $\alpha = 0.5$, and $\alpha = 0.3$.}
	\label{fig:compressible_SNR_vs_rhok_p2}
\end{figure*}

\section{Stylized Applications}\label{sec:applications}
In this section, we apply standard and weighted $\ell_1$ minimization to recover real video and audio signals that are compressively sampled.

\subsection{Recovery of video signals}
One natural application for weighted $\ell_1$ minimization is video compressed sensing. Traditional video acquisition techniques capture a full frame (or image) in the pixel domain at a specific frame rate. The number of pixels acquired per image defines the spatial sampling rate, while the number of frames acquired per second defines the temporal sampling rate. Since the temporal sampling rate is usually high, a group of adjacent video frames are temporally correlated which is reflected in their spatial transform coefficients having nonzero entries in roughly the same locations. 

Our aim here is to reduce the number of samples acquired for each video frame while keeping the same reconstruction quality by recovering using weighted $\ell_1$ minimization. Here, we assume that for every video frame $j$, the measurements $y_j$, $j \in \{0, 1, \hdots, m-1\}$, are acquired by storing the readings of a random subset of the CCD array with $m$ denoting the total number of frames in the video sequence. Let $n_j$ be the number of measurements acquired per frame $j$ and $N$ be the spatial resolution (number of pixels) to be recovered per frame. Let $D$ be the spatial sparsifying transform. The measurement matrix $A_j$ can then be written as $A_j = R_jD$, where $R_j$ is an $n_j \times N$ restriction matrix, and $D$ is an orthonormal basis. Note that the restriction matrix $R_j$ randomly selects $n_j$ pixels from the $N$ pixels in the CCD array to store their readings.

For the first frame, $j = 0$, $n_0$ measurements are captured and the
transform coefficients $x_0$ are recovered by solving the standard
$\ell_1$ minimization problem
\begin{equation}\nonumber
  \hat{x}_0 = \argmin\limits_{x}\ \|x\|_1 \ \text{subject to} \ Ax = y_0.
\end{equation}
For every subsequent frame $j \geq 1$, a support estimate
$\tilde{V}_j$ is chosen to be the union of the locations of the nonzero
entries of $\hat{x}_{j-1}$ and $\hat{x}_{j-2}$ that contribute a
certain percentage of the energy of $\hat{x}_{j-1}$ and
$\hat{x}_{j-2}$, respectively. Consequently, the coefficients
$\hat{x}_j$ are recovered from $n_j < n_0$ measurements $y_j$ by
solving the following weighted $\ell_1$ minimization problem
\begin{equation*}
  \hat{x}_j = \argmin_{x}\ \|x\|_{1,\mathrm{w}} \ \text{subject to} \
  Ax = y_j,
  \quad\text{with}\quad
  w_i=\begin{cases}
    1,       & i \in \tilde{V}_j^c,
  \\\omega,  & i \in \tilde{V}_j,
\end{cases}
\end{equation*}
where $0 \leq \omega \leq 1$.

In our experiments, we use the Foreman sequence at QCIF resolution,
i.e., every frame contains $144\times 176$ pixels. We only consider
the luma (grayscale) component of the sequence. Every frame is split
into four blocks, each of size $N = 72\times 88$ which are processed
independently. We set $n_0 = N/2$ and $n_j = N/2.2$ and $n_j = N/2.4$
for $j \geq 1$. The two dimensional discrete cosine transform (DCT) is used as the spatial
sparsifying basis allowing for the support estimate $\tilde{V}_j$ to
include the DC component and the union of the AC coefficients that
contribute to 97\% of the energy in the AC coefficients of each of
$\hat{x}_{j-1}$ and $\hat{x}_{j-2}$. The signals $\hat{x}_j$ are then
recovered using weighted $\ell_1$ minimization for $\omega$ equal to
0, 0.1, 0.5, and 1.

Figure~\ref{fig:foreman_wl1_alg2} illustrates the recovery of the first
30 frames of the Foreman sequence using weighted $\ell_1$
minimization. The reconstruction quality is reported in terms of the
peak signal to noise ratio (PSNR) given by the expression
\begin{equation}
	\mathrm{PSNR}(x, \hat{x}) = 10\log_{10}\left(\frac{N\times 255^2}{\|x-\hat{x}\|_2^2}\right).
\end{equation} 
The figure demonstrates that recovery with $\omega = 0.5$ results in an improvement in PSNR averaging around 1~dB compared to standard $\ell_1$ using the same number of measurements. A striking observation is that weighted $\ell_1$ minimization outperforms standard $\ell_1$ also with fewer measurements, i.e., in the case where $n_j = n_0$, $\forall j$ for standard $\ell_1$, whereas $n_j = n_0/2.2$ for weighted $\ell_1$. 

\begin{figure*}
	\centering
	\includegraphics[width=7in]{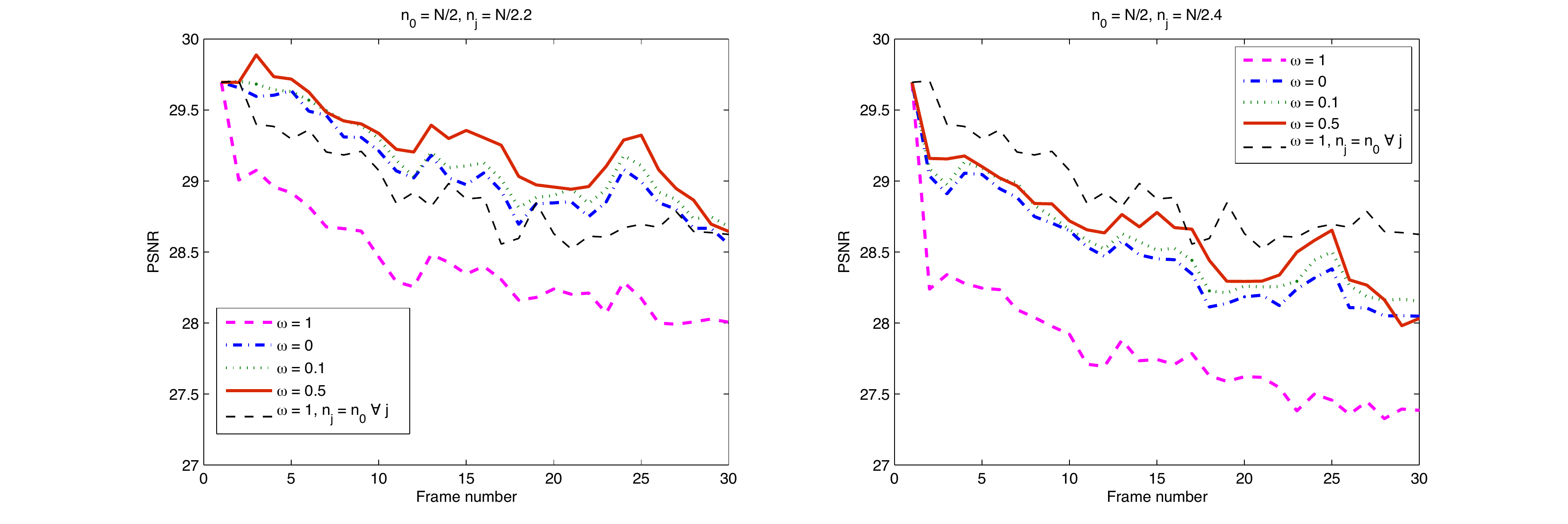}
	\caption{Recovery of the first 30 frames of the Foreman
          sequence at QCIF resolution. The first frame is recovered
          from $n_0 = N/2$ measurements, while the remaining frames
          are recovered from (a) $n_j = N/2.2$ and (b) $n_j = N/2.4$
          measurements. Recovery is performed using weighted $\ell_1$
          minimization with $\omega \in \{0, 0.1, 0.5, 1\}$. The
          support estimate is derived from the union of the supports
          of the previous two frames. The black curve corresponds to
          the recovered PSNR using standard $\ell_1$ minimization with
          a fixed number of measurements $n_j = n_0$, $\forall j \in
          \{1,\dots, 30\}$.}
	\label{fig:foreman_wl1_alg2}
\end{figure*}
	
\subsection{Recovery of audio signals}
For our second stylized application, we examine the performance of
weighted $\ell_1$ minimization for the recovery of compressed sensing
measurements of speech signals. In particular, the original signals
are sampled at $44.1$ kHz, but only $1/4$th of the samples are retained
(with their indices chosen randomly from the uniform
distribution). This yields the measurements $y=Rs$, where $s$ is the
speech signal and $R$ is a restriction (of the identity)
operator. Consequently, by dividing the measurements into blocks of size $N$, we
can write $y=[y_1^T,y_2^T,...]^T$. Here each $y_j=R_j s_j$ is the
measurement vector corresponding to the $j$th block of the signal, and $R_j\in\R^{n_j\times N}$ is the associated restriction matrix. The signals we use in our experiments consist of 21 such blocks.
%
We make the following assumptions about speech signals: 
\begin{enumerate}
\item The signal blocks are compressible in the DCT domain (for example, the MP3
compression standard uses a version of the  DCT to compress audio
signals.)
\item The support set corresponding to the
largest coefficients in adjacent blocks does not change much from
block to block. 
\item Speech signals have large low-frequency coefficients.
\end{enumerate}
Thus, for the reconstruction of the $j$th block, we choose the support estimate  $\widetilde{T}=\widetilde{T}^1\cup\widetilde{T}^2$. Here, $\widetilde{T}^1$ is the set corresponding to frequencies up to 4kHz and $\widetilde{T}^2$
is the set corresponding to the largest $n_j/16$ recovered coefficients of the
previous block (for the first block $\widetilde{T}^2$ is empty).  The results of experiments on two speech signals (one male and one female) with
$N=2048$, and $\omega \in \{0,1/6,2/6,\ldots,1\}$ are illustrated in
Figure~\ref{fig:audio}.
\begin{figure*}[h]
	\centering
	\includegraphics[width=4in]{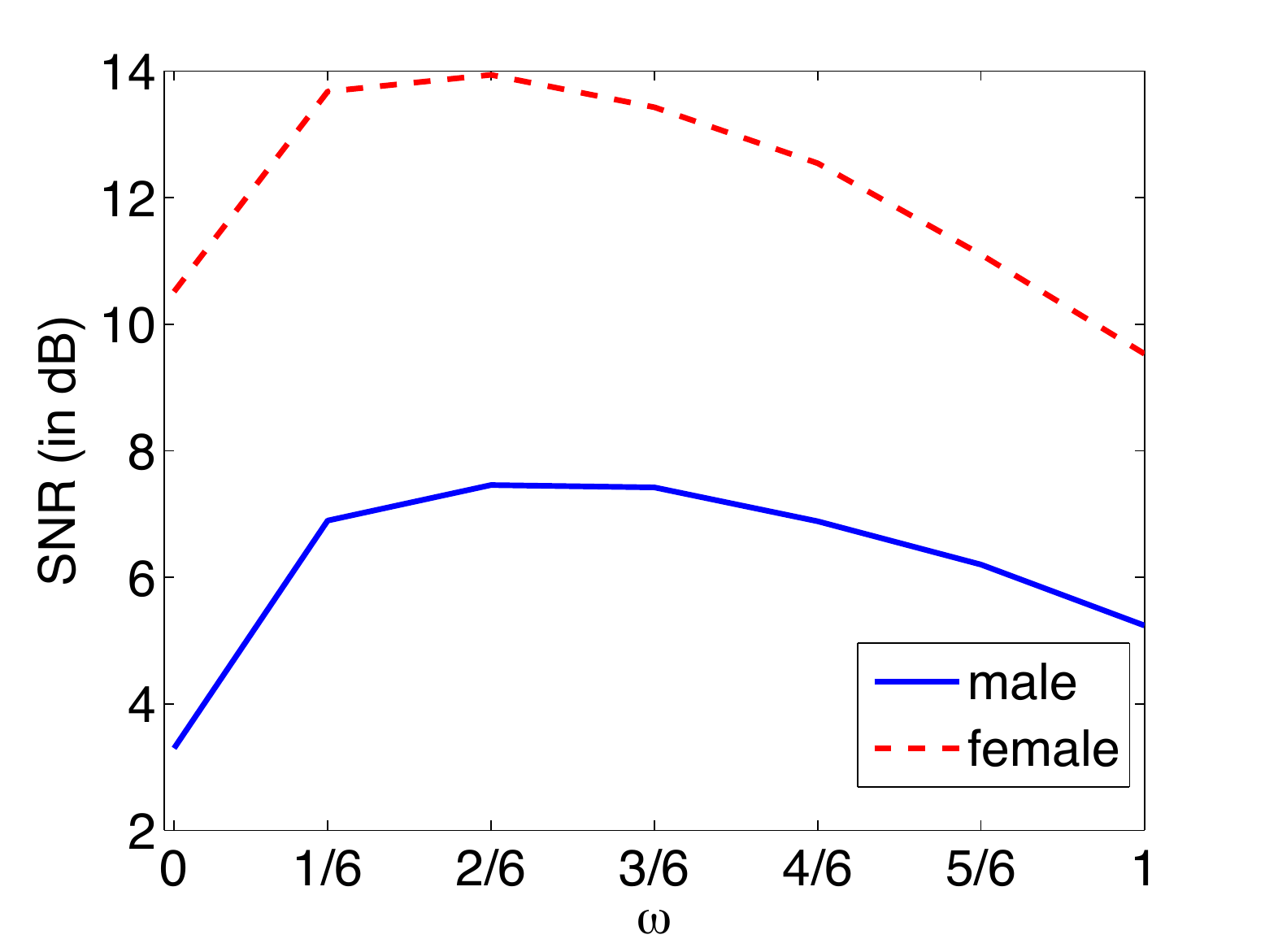}
	\caption{SNRs of two reconstructed signals (male and female
          voices) from compressed sensing measurements plotted against $\omega$. For both speech signals, an intermediate value of $\omega$ yields the best performance.}
	\label{fig:audio}
\end{figure*}


\section{Proof of Theorem \ref{thm:weighted_L1_recovery}}\label{sec:Proof}
 Recall that $\widetilde{T}$, an arbitrary subset of $\{1,2,\ldots,N\}$,  is of size $\rho k$ where $0 \leq \rho \leq a$ and $a$ is some number larger than 1. Let the set $\widetilde{T}_{\alpha} = T_0 \cap \widetilde{T}$ and $\widetilde{T}_{\beta} = T_0^c \cap \widetilde{T}$, where $|\widetilde{T}_{\alpha}| = \alpha |\widetilde{T}| = \alpha \rho k$ and $\alpha + \beta = 1$.
Figure \ref{fig:Sets_Weights} illustrates these sets and shows the relationship to the weight vector $\mathrm{w}$.
\begin{figure}[h]
	\centering
		\includegraphics[width=5in]{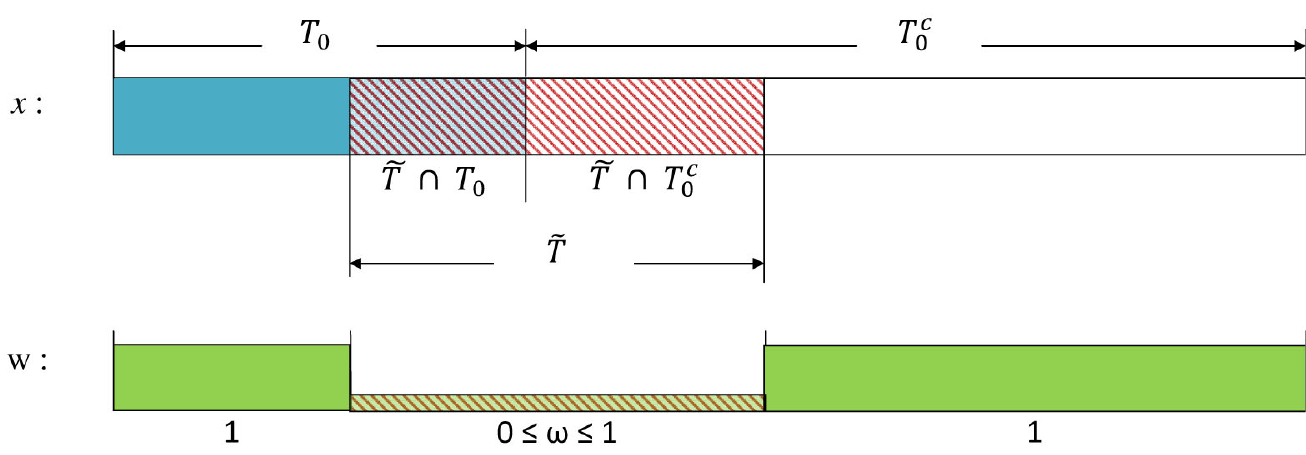}
	\caption{Illustration of the signal $x$ and weight vector $\mathrm{w}$ emphasizing the relationship between the sets $T_0$ and $\widetilde{T}$.  }
	\label{fig:Sets_Weights}
\end{figure}


Let $x^* = x + h$  be a minimizer of the weighted $\ell_1$ problem  (\ref{eq:weighted_L1}). Then
\begin{equation*}
	 \|x + h\|_{1,\mathrm{w}}  \leq \|x\|_{1,\mathrm{w}}.
\end{equation*}
Moreover, by the choice of weights in \eqref{eq:weighted_L1}, we have
\begin{equation*}
 \omega\|x_{\widetilde{T}} + h_{\widetilde{T}}\|_1 + \|x_{\widetilde{T}^c} + h_{\widetilde{T}^c}\|_1 \leq \omega\|x_{\widetilde{T}}\|_1 + \|x_{\widetilde{T}^c}\|_1. 
\end{equation*}
Consequently, 
\begin{equation*}
\begin{array}{ll}
 & \|x_{\widetilde{T}^c \cap T_0} + h_{\widetilde{T}^c \cap T_0}\|_1 + \|x_{\widetilde{T}^c \cap T_0^c} + h_{\widetilde{T}^c \cap T_0^c}\|_1 
	  \quad+ \quad \omega \|x_{\widetilde{T} \cap T_0} + h_{\widetilde{T} \cap T_0}\|_1 + \omega\|x_{\widetilde{T} \cap T_0^c} + h_{\widetilde{T} \cap T_0^c}\|_1 \\ 
	& \quad \quad \quad \quad \quad \quad \quad \leq \quad \|x_{\widetilde{T}^c \cap T_0}\|_1 + \|x_{\widetilde{T}^c \cap T_0^c}\|_1 + \omega\|x_{\widetilde{T} \cap T_0}\|_1 + \omega\|x_{\widetilde{T} \cap T_0^c}\|_1.
\end{array}
\end{equation*}
Next, we use the forward and reverse triangle inequalities to get
\begin{equation}\nonumber
	\omega \|h_{\widetilde{T} \cap T_0^c}\|_1 + \|h_{\widetilde{T}^c \cap T_0^c}\|_1  \leq \|h_{\widetilde{T}^c \cap T_0}\|_1 + \omega \|h_{\widetilde{T} \cap T_0}\|_1  + 2\left(\|x_{\widetilde{T}^c \cap T_0^c}\|_1 + \omega \|x_{\widetilde{T} \cap T_0^c}\|_1\right).
\end{equation}
Adding and subtracting $\omega\|h_{\widetilde{T}^c \cap T_0^c}\|_1$ on the left hand side, and $\omega\|h_{\widetilde{T}^c \cap T_0}\|_1$ on the right, we obtain
\begin{equation}\nonumber
\begin{array}{ll}
\omega \|h_{\widetilde{T} \cap T_0^c}\|_1 + \omega\|h_{\widetilde{T}^c \cap T_0^c}\|_1 & + \quad \|h_{\widetilde{T}^c \cap T_0^c}\|_1  -\omega \|h_{\widetilde{T}^c \cap T_0^c}\|_1\\
	&\leq \omega \|h_{\widetilde{T} \cap T_0}\|_1 + \omega \|h_{\widetilde{T}^c \cap T_0}\|_1 +  \|h_{\widetilde{T}^c \cap T_0}\|_1 - \omega\|h_{\widetilde{T}^c \cap T_0}\|_1 \\
	 & \quad \quad + \quad 2\left(\omega \|x_{\widetilde{T} \cap T_0^c}\|_1 + \omega\|x_{\widetilde{T}^c \cap T_0^c}\|_1 + \|x_{\widetilde{T}^c \cap T_0^c}\|_1 - \omega\|x_{\widetilde{T}^c \cap T_0^c}\|_1\right).
\end{array}
\end{equation}
Since $\|h_{T_0^c}\|_1 =  \|h_{\widetilde{T} \cap T_0^c}\|_1 + \|h_{\widetilde{T}^c \cap T_0^c}\|_1$, this easily reduces to 
\begin{equation}\label{eq:weighted_optimality}
	\omega\|h_{T_0^c}\|_1 +  (1-\omega)\|h_{\widetilde{T}^c \cap T_0^c}\|_1 
	 \leq  \omega\|h_{T_0}\|_1 + (1-\omega)\|h_{\widetilde{T}^c \cap T_0}\|_1 + 2\left(\omega\|x_{T_0^c}\|_1 + (1-\omega)\|x_{\widetilde{T}^c \cap T_0^c}\|_1\right).
\end{equation}
But, we can also write 
\begin{equation}\nonumber
\|h_{T_0^c}\|_1
 = \omega\|h_{T_0^c}\|_1 + (1-\omega)\|h_{\widetilde{T}^c \cap T_0^c}\|_1 + (1-\omega)\|h_{\widetilde{T} \cap T_0^c}\|_1.
\end{equation}
Combining the above with \eqref{eq:weighted_optimality}, we obtain
\begin{equation}\nonumber
\begin{array}{lrl}
\Rightarrow & \|h_{T_0^c}\|_1 & \leq \omega\|h_{T_0}\|_1 + (1-\omega)\|h_{\widetilde{T}^c \cap T_0}\|_1 + (1-\omega)\|h_{\widetilde{T} \cap T_0^c}\|_1 + 2\left(\omega\|x_{T_0^c}\|_1 + (1-\omega)\|x_{\widetilde{T}^c \cap T_0^c}\|_1\right)\\
&& \quad = \omega\|h_{T_0}\|_1 + (1-\omega)\left(\|h_{\widetilde{T}^c \cap T_0}\|_1 + \|h_{\widetilde{T} \cap T_0^c}\|_1\right) + 2\left(\omega\|x_{T_0^c}\|_1 + (1-\omega)\|x_{\widetilde{T}^c \cap T_0^c}\|_1\right).
\end{array}
\end{equation}
Since, the set $\widetilde{T}_{\alpha} = T_0 \cap \widetilde{T}$, we can write $\|h_{\widetilde{T}^c \cap T_0}\|_1 + \|h_{\widetilde{T} \cap T_0^c}\|_1 = \|h_{T_0 \cup \widetilde{T}\setminus \widetilde{T}_{\alpha}}\|_1$ and simplify the bound on $\|h_{T_0^c}\|_1$ to the following expression:
\begin{equation}\label{eq:weighted_hT0c1}
	\|h_{T_0^c}\|_1 \leq \omega\|h_{T_0}\|_1 + (1-\omega)\|h_{T_0 \cup \widetilde{T}\setminus \widetilde{T}_{\alpha}}\|_1 + 2\left(\omega\|x_{T_0^c}\|_1 + (1-\omega)\|x_{\widetilde{T}^c \cap T_0^c}\|_1\right).
\end{equation}

Next we sort the coefficients of $h_{T_0^c}$ partitioning $T_0^c$ it into disjoint sets $T_j, j \in \{1,2,\ldots\}$ each of size $ak$, where $a > 1$. That is, $T_1$ indexes the $ak$ largest in magnitude coefficients of $h_{T_0^c}$, $T_2$ indexes the second $ak$ largest in magnitude coefficients of $h_{T_0^c}$, and so on. Note that this gives  $h_{T_0^c} = \sum_{j \geq 1} h_{T_j}$, with
\begin{equation}\label{eq:hTj2}
	\|h_{T_j}\|_2 \leq \sqrt{ak} \|h_{T_j}\|_{\infty} \leq (ak)^{-1/2} \|h_{T_{j-1}}\|_1.
\end{equation}
Let $T_{01} = T_0 \cup T_1$, then using (\ref{eq:hTj2}) and the triangle inequality we have
\begin{equation}\label{eq:hT01c2_sum_bound}
\begin{array}{lrl}
	&\|h_{T_{01}^c}\|_2 & \leq \sum\limits_{j \geq 2} \|h_{T_j}\|_2  \leq (ak)^{-1/2} \sum\limits_{j \geq 1} \|h_{T_j}\|_1\\
	&& \leq (ak)^{-1/2} \|h_{T_0^c}\|_1.
\end{array}
\end{equation}
Combining the above expression with (\ref{eq:weighted_hT0c1}) we get
\begin{equation}\label{eq:weighted_hT01c2}
	\|h_{T_{01}^c}\|_2 \leq (ak)^{-1/2} \left(\omega\|h_{T_0}\|_1 + (1-\omega)\|h_{T_0 \cup \widetilde{T}\setminus \widetilde{T}_{\alpha}}\|_1 + 2\left(\omega\|x_{T_0^c}\|_1 + (1-\omega)\|x_{\widetilde{T}^c \cap T_0^c}\|_1\right) \right).
\end{equation}

Next, consider the feasibility of $x^*$ and $x$. Both vectors are feasible, so we have $\|Ah\|_2 \leq 2\epsilon$ and
\begin{equation}\nonumber
\begin{array}{lrl}
 \|Ah_{T_{01}}\|_2 &\leq& 2\epsilon + \|Ah_{T_{01}^c}\|_2 
\leq 2\epsilon + \sum\limits_{j \geq 2} \|Ah_{T_j}\|_2 \\
& \leq& 2\epsilon + \sqrt{1+\delta_{ak}} \sum\limits_{j \geq 2} \|h_{T_j}\|_2.
\end{array}
\end{equation}
From (\ref{eq:hT01c2_sum_bound}) and (\ref{eq:weighted_hT01c2}) we get
\begin{equation}\nonumber
\begin{array}{lrl}
& \|Ah_{T_{01}}\|_2 &\leq 2\epsilon + 2\frac{\sqrt{1+\delta_{ak}}}{\sqrt{ak}} \left(\omega\|x_{T_0^c}\|_1 + (1-\omega)\|x_{\widetilde{T}^c \cap T_0^c}\|_1\right)\\
&& \quad + \quad \omega\frac{\sqrt{1+\delta_{ak}}}{\sqrt{ak}}\|h_{T_0}\|_1 + (1-\omega)\frac{\sqrt{1+\delta_{ak}}}{\sqrt{ak}}\|h_{T_0 \cup \widetilde{T}\setminus \widetilde{T}_{\alpha}}\|_1 
\end{array}
\end{equation}
Noting that $|T_0 \cup \widetilde{T}\setminus \widetilde{T}_{\alpha}| = (1 + \rho - 2\alpha\rho)k$,
\begin{equation}\nonumber
\begin{array}{lrl}
& \sqrt{1-\delta_{(a+1)k}}\|h_{T_{01}}\|_2 &\leq 2\epsilon + 2\frac{\sqrt{1+\delta_{ak}}}{\sqrt{ak}} \left(\omega\|x_{T_0^c}\|_1 + (1-\omega)\|x_{\widetilde{T}^c \cap T_0^c}\|_1\right)\\
&& \quad + \quad \omega\frac{\sqrt{1+\delta_{ak}}}{\sqrt{a}}\|h_{T_0}\|_2 + (1-\omega)\frac{\sqrt{1+\delta_{ak}}}{\sqrt{a}}\sqrt{1 + \rho - 2\alpha\rho}\|h_{T_0 \cup \widetilde{T}\setminus \widetilde{T}_{\alpha}}\|_2.
\end{array}
\end{equation}
Since the set $T_1$ contains the largest $ak$ coefficients of
$h_{T_0^c}$ with $a > 1$, and $|\widetilde{T}\setminus
\widetilde{T}_{\alpha}| = (1 - \alpha)\rho k \le ak$, then $\|h_{T_0
  \cup \widetilde{T}\setminus \widetilde{T}_{\alpha}}\|_2 \leq
\|h_{T_{01}}\|_2$. We also have $\|h_{T_0}\|_2 \leq \|h_{T_{01}}\|_2$,
thus
\begin{equation}\label{eq:weighted_hT01_2}
	\|h_{T_{01}}\|_2 \leq \frac{2\epsilon + 2\frac{\sqrt{1+\delta_{ak}}}{\sqrt{ak}} \left(\omega\|x_{T_0^c}\|_1 + (1-\omega)\|x_{\widetilde{T}^c \cap T_0^c}\|_1\right)}{\sqrt{1-\delta_{(a+1)k}} - \frac{\omega + (1-\omega) \sqrt{1+\rho-2\alpha\rho}}{\sqrt{a}}\sqrt{1+\delta_{ak}}}.
\end{equation}

Finally, using $\|h\|_2 \leq \|h_{T_{01}}\|_2 + \|h_{T_{01}^c}\|_2$, we combine (\ref{eq:weighted_hT01c2}) and (\ref{eq:weighted_hT01_2}) to get
\begin{equation}\label{eq:weighted_h2}
	\|h\|_2 \leq \frac{2\left(1+\frac{\omega + (1-\omega)\sqrt{1+\rho-2\alpha\rho}}{\sqrt{a}}\right)\epsilon + 2 \frac{
\sqrt{1-\delta_{(a+1)k}} + \sqrt{1+\delta_{ak}}}{\sqrt{ak}}\left(\omega\|x_{T_0^c}\|_1 + (1-\omega)\|x_{\widetilde{T}^c \cap T_0^c}\|_1\right) }{\sqrt{1-\delta_{(a+1)k}} - \frac{\omega + (1-\omega)\sqrt{1+\rho-2\alpha\rho}}{\sqrt{a}}\sqrt{1+\delta_{ak}}  },
\end{equation}
with the condition that the denominator is positive, equivalently 
\begin{equation}\label{eq:weighted_RIP_relationship}
\delta_{ak} + \frac{a}{\left(\omega + (1-\omega)\sqrt{1+\rho-2\alpha\rho}\right)^2}\delta_{(a+1)k} < \frac{a}{\left(\omega + (1-\omega)\sqrt{1+\rho-2\alpha\rho}\right)^2} - 1. 
\end{equation}
\qed

\section{Acknowledgment}
The authors would like to thank 
the anonymous reviewers for their positive and constructive feedback. Their input has helped improve the presentation of the paper and has made our coverage of the topic more comprehensive. 

\bibliographystyle{IEEEtran}
\bibliography{sparse}

\end{document}